\documentclass[3p]{elsarticle}
\usepackage{amsfonts}
\usepackage{amsmath}
\usepackage{algorithm,algorithmicx,algpseudocode}
\usepackage{subfig}
\usepackage{multirow}
\usepackage{url}
\usepackage{graphicx}
\usepackage{caption}

\usepackage{array}
\usepackage{color}
\usepackage{colortbl}

\DeclareGraphicsExtensions{.pdf,.png,.jpg,.eps}

\usepackage[]{changes}
\definechangesauthor[color=red]{AP}
\definechangesauthor[color=blue]{EB}
\definechangesauthor[color=green]{RB}

\usepackage{soul}

\sethlcolor{yellow}

\usepackage[]{todonotes}

\begin{document}
\begin{frontmatter}

\title{Making opportunistic networks in IoT environments CCN-ready:\\
	a performance evaluation of the MobCCN protocol}

\author[cnr]{Eleonora Borgia} \ead{eleonora.borgia@iit.cnr.it}
\author[cnr]{Raffaele Bruno} \ead{raffaele.bruno@iit.cnr.it}
\author[cnr]{Andrea Passarella\corref{cor1}} \ead{andrea.passarella@iit.cnr.it}
\cortext[cor1]{Corresponding author}
\address[cnr]{Institute for Informatics and Telematics, National
Research Council. Via G. Moruzzi 1, 56214 Pisa, Italy}

\begin{abstract}
In future IoT environments it is expected that the role of personal devices of mobile users in the physical area where IoT devices are deployed will become more and more important. In particular, due to the push towards decentralisation of services towards the edge, it is likely that a significant share of data generated by IoT devices will be needed by other (mobile) nodes nearby, while global Internet access will be limited only to a small fraction of data. In this context, opportunistic networking schemes can be adopted to build efficient content-centric protocols, through which data generated by IoT devices (or by mobile nodes themselves) can be accessed by the other nodes nearby. In this paper, we propose MobCCN, which is an ICN-compliant protocol for this heterogeneous environment. MobCCN is designed to implement the routing and forwarding mechanisms of the main ICN realisations, such as CCN. The original aspect of MobCCN is to implement an efficient opportunistic networking routing scheme to populate the Forwarding Interest Base (FIB) tables of the nodes, in order to guide the propagation of Interest packets towards nodes that store the required data. Specifically, MobCCN defines the \emph{utility} of each node as a forwarder of Interest packets for a certain type of content, such that Interest packets can be propagated along a positive utility gradient, until reaching some node storing the data. We evaluate MobCCN against protocols representing two possible endpoints of the spectrum, respectively in terms of minimising the data delivery delay and the resource consumption. Performance results show that MobCCN is very effective and efficient, as it guarantees very high delivery rates and low delays, while keeping the total generated traffic at a reasonable level and also saving local resources.
\end{abstract}


\end{frontmatter}

\section{Introduction}
\label{sec:intro}

Internet of Things (IoT) and opportunistic networks have not often being correlated with each other and considered as part of a unique networking environment. However, we believe that they are significantly complementary, and using opportunistic networks in IoT environment can result in very efficient and dynamic IoT systems deployments.

In this paper we consider such a mixed network environment. Concerning IoT, we assume that a large number of fixed IoT devices are spread in a given physical area. As far as the opportunistic network, we consider that a number of mobile users equipped with personal devices (e.g., smartphones or tablets) move in the same physical area. Both types of devices can generate and consume data. IoT devices can be sensors and/or actuators, which thus generate data about the physical environment and request data to decide how to act upon it. Users devices request data based on the applications running on them, but can also act as mobile sensors, thus collecting data that are required by the other nodes in the network.

We also assume that data generated by these devices are primarily of interest for other nodes in the same physical area.  While the protocol we propose can support also data access from remote locations over the Internet, the primary focus of the paper is on accessing data from the same physical area where it is generated. 
A use-case example for the scenario envisioned is the "Smart Campus", which refers to a delimited physical area composed of several buildings where a number of different IoT sensors and/or actuators are located and monitor the physical environment. Then, professors, students, technicians, users in general enrich the scenario moving freely within the area, and using their devices to access data directly and/or to forward it. Data generated by such IoT devices is mainly of local value and of interests for users spend time (everyday or occasionally) in that area. For example, as part of smart access services to the campus resources, students might be interested in the status of a given physical space (e.g., the number of people in the canteen or the cafeteria) or at the existence of scientific events in the area. Technicians might be interested in statistically understanding student mobility to improve intra-campus travel modes. IoT devices generate the above contents in the form of simple text information or more complex data such as pictures or short videos, and users with their movements make them available to those who have requested them. Similar examples might be envisioned in different contexts such as a neighbourhood or a portion of the city (e.g., a park). In these cases, users would be interested in the content itself more than in the specific device generating it, which justifies the use of a data-centric network approach. Moreover, using mobile nodes as data carries allows to use low-energy communication technologies for IoT devices (such as Bluetooth Low Energy~\cite{BLE:2010}), and to avoid unnecessary traffic load on cellular infrastructures, which might be a severe issue also for the latest cellular technologies, as discussed next. Our approach is thus aligned with the trend of decentralization of functions at the (extreme) edge of the networks~\cite{Lopez:2015aa}, which is particularly relevant for IoT applications~\cite{Borgia20141}.

In this context, the paper presents and evaluates MobCCN, which is a protocol for accessing data generated by IoT devices, in presence of mobile opportunistic networks. MobCCN is compliant with the Information-Centric Networking (ICN) principles. The ICN paradigm~\cite{6563278} has been initially conceived for fixed infrastructure-based networks, to turn them into native content-centric networks. It has then been proposed also for mobile networking environments and IoT in particular~\cite{7437030,Baccelli:2014:ICN:2660129.2660144}, due to the fact that, for a very significant set of IoT applications, it is important to acquire data generated by devices, while accessing - and, thus, addressing - each and every individual device is less important.  

Existing ICN-inspired solutions for IoT and mobile networks are not well suited for the network environment envisioned. The standard approach uses IoT devices as mere producers of raw (sensing) data, fixed gateways located at the boundary between the wireless/mobile and the fixed network to collect data, and ICN mechanisms to access data stored at the gateways from anywhere in the Internet. This approach has several limitations: (i) it assumes a wireless infrastructure through which IoT devices can always communicate with gateways. Even the latest cellular technologies might not provide sufficient capacity to support all IoT applications (in addition to the rest of non-IoT applications), considering the predicted evolution of traffic demand vs. cellular capacity~\cite{cisco16}; (ii) approaches such as extreme densification of cellular networks~\cite{6171992,6824752} might prove too costly, as the infrastructure might need too frequent updates to cope with traffic demands; (iii) \emph{local} data consumption patterns, whereby data is mostly accessed in the vicinity where it is generated, as well as mobile nodes in the areas where the data is generated are not exploited, and (iv) most of the protocols for transferring data between the sensors and the gateways are not ICN-compliant, or where implemented, they focus on different aspects as we discuss in Section~\ref{sec:related}.Therefore, in this paper we propose an approach where ICN data management functions are entirely implemented in the mobile network formed by the IoT nodes.

Our proposed protocol, MobCCN, is a data management solution compliant with the CCN/NDN mechanisms~\cite{Jacobson:2009:NNC:1658939.1658941,Zhang:2014:NDN:2656877.2656887}. CCN was used as one of the reference ICN implementations, although MobCCN algorithms could be used with other ICN architectures (e.g., NetInf~\cite{Dannewitz2013721}, DONA~\cite{Koponen:2007:DNA:1282380.1282402}, PURSUIT~\cite{6231280}, \ldots) and easily extend to also support remote access to data generated by IoT devices. MobCCN exploits the presence of mobile devices forming an opportunistic network to provide data-centric access to data generated by IoT devices or mobile nodes themselves. This is achieved by computing, at each mobile node, a \emph{utility} value for each type of data available in the network (possibly generated by IoT devices or by mobile nodes). The utility $U_{p,i}$ of mobile node $i$ for content type $p$ is defined as the utility of $i$ to access data of type $p$, and consists in a \emph{direct} utility value, which is computed for the content types available locally or available on the IoT devices/other mobile nodes they encounter directly, and an \emph{indirect} utility, which is a function of the frequency with which they encounter other mobile nodes that can ``provide access'' to the given data type. Obviously, the higher $U_{p,i}$, the more $p$ is ``useful'' to access data of type $p$. 
The use of utility creates a distributed index of the data types available in the network, together with a quantitative value that can be used to appropriately direct data access requests towards the appropriate nodes. Specifically, when a data request is generated, it is opportunistically forwarded among mobile nodes based on a positive gradient defined by the utility values, until some of the devices providing the data is reached. Then, according to the standard CCN mechanisms, data follows the reverse path established by the data request. In CCN terminology (see Section~\ref{sec:ccn}), the gradient is used to forward the Interest packets, and the standard breadcrumbs mechanism is used to forward the corresponding Data packets.

We have implemented a MobCCN prototype in the CCN-lite~\cite{ccn-lite} codebase, which is a reference implementation of CCN for resource-constrained devices. Using the OMNET++ simulation environment, we have evaluated its performance against reference alternative protocols from the literature. Specifically, MobCCN is compared against two ``extreme'' cases. On the one hand, we consider a protocol where both Interest and Data packets are sent via Epidemic forwarding~\cite{vahdat2000epidemic}. As explained in detail in Section~\ref{sec:peva-settings}, this allows us to test MobCCN against a protocol that achieves the minimum delay for receiving the requested data. As expected, end-to-end delay is higher in MobCCN than in Epidemic. However, the overhead in terms of network traffic is one order of magnitude less in MobCCN than in Epidemic. This shows that, contrarily to what often advocated in the literature about ICN for IoT, disseminating content availability - as in the original ICN design - is useful also in mobile networking environments in terms of resource consumption. Then, we consider an opposite version of Epidemic, whereby only one copy of Interest packets is allowed in the network, and each intermediate node forwards it to a randomly chosen neighbour. Data packets are sent back through the breadcrumbs established by the propagation of the Interest packet. This is the least resource consuming version of an Epidemic-like protocol, as neither Interest nor Data packets are flooded. We show that MobCCN significantly outperforms this version of Epidemic in terms of delivery rate and delay, with an acceptable additional cost in terms of resource consumption. Finally, we compare MobCCN and this version of Epidemic in the cases when caching of Data packets at nodes is enabled or not, respectively. Our results show that, while the performance in terms of end-to-end delay and delivery rate of MobCCN does not significantly drop when caching is disabled, the Epidemic-based protocol suffers significantly.  This is a further confirmation of the advantage of the use of the MobCCN utility-based mechanism: in MobCCN, Interest packets are propagated along appropriate opportunistic paths towards nodes storing the requested data, and thus it remains effective also without caching. The analysis presented in Sections~\ref{sub:mobcnn-vs-ideal-epidemic} and~\ref{sub:comparison-epidemic-ccn} refers to an homogeneous mobility environment, i.e., where nodes encounter each other according to identical stochastic processes. In cases of heterogeneous mobility, the advantage of MobCCN over Epidemic-style protocols is even higher. Conversely, to achieve the same MobCCN performance, Epidemic-style protocols have to generate huge network traffic or to excessively waste the local resources of the device.
All in all, therefore, with respect to the existing literature, our results show that (i) data-centric (CCN-compliant) access can be efficiently supported in IoT environments by exploiting mobile nodes forming opportunistic networks, thus, without necessarily relying on storing all data in fixed gateways; (ii) utility-based opportunistic protocols used to propagate Interest packets are a viable approach to effective and efficient CCN-compliant protocols in mobile networking environments.

The rest of the paper is organised as follows. Section~\ref{sec:related} describes the related literature. Section~\ref{sec:ccn} provides, as background material, a concise presentation of the main CCN mechanisms. In Section~\ref{sec:mobccn} we describe the MobCCN algorithms. Section~\ref{sec:peva-settings} describes in detail the performance evaluation strategy, while Section~\ref{sec:results} presents the performance results. Finally, Section~\ref{sec:conclusions} draws the main conclusions of the paper.

\section{Related work}
\label{sec:related}
Due to the data-centric nature of IoT applications, applying ICN to IoT is a hot
research topic. A first class of proposals
(e.g., ~\cite{lindgren-icnrg-efficientiot-03,Amadeo:2014:MDR:2660129.2660148,Biswas:2013:CIH:2534169.2491691,6815203,7444734})
assume that all IoT data is transferred to fixed collection points at the edge
of the network. They design ICN protocols running on the \emph{fixed} network
\emph{only} to adapt ICN mechanisms to the specificities of IoT applications in
generating and consuming data. For example, emphasis is given on how to support,
with standard ICN mechanisms, access to streams of sensor data or to one shot
sensor readings. Thus, IoT and mobile devices at the edge of the network are
left out of the ICN architecture.

Another class of papers implement ICN functions on IoT devices,
e.g., ~\cite{Baccelli:2014:ICN:2660129.2660144,6686486,6970725}. However,
most of these studies focus only on fixed devices such as tiny sensor nodes, and
therefore the main target is the efficient implementation of ICN mechanisms in
terms of energy and memory resource consumption. Then, ICN is typically seen as a
replacement for sensor-oriented protocol stacks such as IETF CoAP/RPL, and
performance is assessed in terms of resource footprint on small IoT devices.

In other cases, the presence of mobile devices is taken into consideration.
Interested users are referred to~\cite{Amadeo2015148} for a survey on routing
and forwarding techniques in these cases. Most of the time, no FIB state is
maintained, as this is assumed to be too resource consuming. However, as no FIB
are used, nodes are necessarily forced to flood Interest packets upon a data
request. To reduce the amount of flooding, one-hop, time-out based suppression
techniques are used, which is a standard technique in the mobile networking
literature \cite{6504249,Angius:2012:BBF:2248361.2248369,6724509,Hail2015}. However,
this is completely agnostic to where data is actually located. On the contrary,
MobCCN implements a lightweight protocol for FIB construction, which allow nodes
to appropriately forward Interest packets based on utility gradients computed
for each data type. In other cases, flooding reduction is aware of the requested
data type (see \cite{Amadeo20141} for examples). However, in these cases
solutions are typically based on MANET-like routing schemes. This is a problem,
because (i) contrary to opportunistic routing, MANET routing has shown to be not
able to support mobility of users~\cite{DBLP:journals/cm/ContiG14}; and (ii) these
solutions are typically not ICN-compliant. MobCCN addresses both these
limitations. Other papers (e.g.,~\cite{7104902}) consider caching policies for ICN
in IoT networks, assuming network environments similar to ours. Caching is
clearly a key issue, which is however orthogonal to MobCCN. Specifically, MobCCN
could easily incorporate any caching policy, as its routing mechanisms would
immediately reflect availability of data on more or less nodes.

Finally, adoption of opportunistic/DTN networking in ICN has been recently
proposed in~\cite{Trossen:2016:TIC:2875951.2875959,7158165}. The RIFE
architecture~\cite{Trossen:2016:TIC:2875951.2875959} sees DTN as one of the
networking transports underlying ICN, primarily to support data access in remote
areas. On the other hand, MobCCN proposes a concrete routing/forwarding scheme
integrating ICN and opportunistic networks. Integrating MobCCN in the RIFE
architecture as one concrete instance of the RIFE approach is an interesting
subject of future investigation. In Navigo~\cite{7158165} authors use a naming
convention to associate geoIDs to content names, and assume a V2V environment,
such that forwarding of interests occur based on the street topology towards the
physical location where data resides. Moreover, Navigo does not use a routing
protocol, and therefore Interests need to be flooded when no information about
the physical location of the destination is known. MobCCN does not assume a
vehicular environment, and therefore its routing and forwarding mechanisms are
more general. Moreover, MobCCN implements a lightweight routing mechanism to
avoid flooding of Interests.

MobCCN takes inspiration from the literature on utility-based routing protocols for opportunistic networks, such as PRoPHET \cite{Lindgren:2003:PRI:961268.961272}, for selecting the most suitable forwarder for Interest packets. Similar to these approaches, it applies the general idea of computing the utility function whenever nodes get in touch. However, differently from them, the MobCCN utility function reflects the appropriateness of encountering a specific content rather than the probability of meeting a node. Indeed, in PRoPHET nodes that are often encountered have a high delivery predictability and thus are considered as the most suitable forwarders. This does not hold in MobCCN, where, as explained in detail in Section~\ref{sec:mobccn}, the utility relies also on the utility of the encountered node towards the requested content. 

This paper extends our prior work in~\cite{Borgia:2016:MCP:2979683.2979695}. Specifically, in this paper we describe in more details the MobCCN algorithms and the strategy behind the performance evaluation of the proposed protocol. We also provide two completely new sets of performance results, complementary with respect to the ones presented in~\cite{Borgia:2016:MCP:2979683.2979695}. Specifically, while in~\cite{Borgia:2016:MCP:2979683.2979695} we compare MobCCN against a protocol providing a lower bound in terms of end-to-end delay, in this paper we also compare MobCCN with respect to two different protocols, which represent lower bounds (in the family of Epidemic-based protocols) for what concerns the use of network resources. We also study the protocol performance with different node density and mobility patterns, and complement the network resource analysis with the study of the device resource. Finally, we discuss in more detail the interplay between IoT environments and opportunistic networking for mobile nodes.

\section{Background: the CCN architecture}
\label{sec:ccn}
Among the different content centric systems that are proposed in the past years,
CCN - and its successor Named Data Networking
(NDN)~\cite{Zhang:2014:NDN:2656877.2656887} - is one of the most popular and
fully-fledged architecture. The communication among CCN nodes relies on the
exchange of two different packets, i.e. \emph{Interest} packets, carrying requests
for contents, and \emph{Data} packets, carrying the content requested.
At each node, CCN maintains three main data structures. The Content Store (CS) acts as a
data cache and stores Data packets forwarded by the node. The Pending Interest
Table (PIT), stores content requests (i.e., Interest packets) that have arrived
at the node, and have not been satisfied yet. The Forwarding Information Base
(FIB) tells how to route Interest packets towards the required content
(analogous to an IP forwarding table). 

The main mechanism on which CCN routing is built is as follows. When a new data is
available at a node, the node advertises its availability on all outgoing
interfaces (or ``faces'' in CCN terminology). Note that CCN is designed for fixed Internet environments, 
thus the concept of interface is well defined. This information is recursively propagated by upstream nodes, i.e., direct neighbours on all interfaces but the one from which the advertisement came, and this
establishes the entries in the FIB for reaching the data. Specifically, as
CCN data names are hierarchical, advertisements are not propagated anymore when
they reach some node that has already advertised data for a name prefix that
matches that of the new data.

\begin{figure}[htpb]
	\centering
	\includegraphics[trim={0, 0, 0, 0}, clip, scale=0.3]{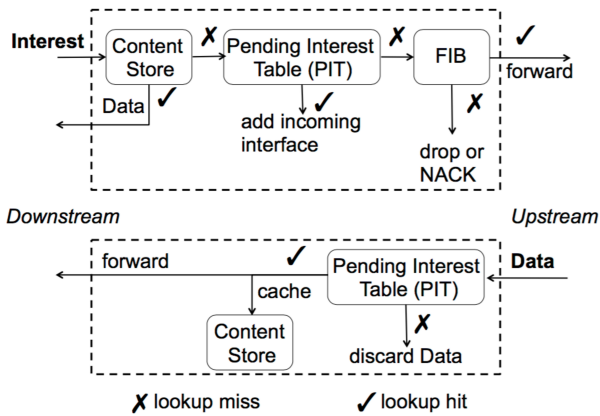}
	\caption{CCN forwarding~\cite{Zhang:2014:NDN:2656877.2656887}}
	\vspace{-0.3cm}
	\label{fig:CCN-forwarding}
\end{figure}

On the other hand, forwarding occurs as follows (see
Figure~\ref{fig:CCN-forwarding}
from~\cite{Zhang:2014:NDN:2656877.2656887}). When a node generates an
Interest packet, this is forwarded from node to node according to the
information stored in the FIBs. The forwarding process ends when
the Interest packet reaches: (i) the source of the required content, or (ii) a
node storing it in its CS. In addition, Interest packets install a reverse route
(breadcrumbs) while being forwarded, which is used to send the data back to the
requesting node. At each intermediate node, a caching policy decides whether
data should be stored or not in the CS, and how to replace existing items, if
the case.

\section{MobCCN}
\label{sec:mobccn}

In order to extend the ICN concept beyond the edge of the fixed network and to
include mobile devices and sensors, some modifications to the standard ICN
routing and forwarding algorithms are needed to cope with the totally different networking environment.
However, it is worth noting that, in order to be fully compatible with the
standard CCN protocol, the basic CCN structures and procedures are maintained in MobCCN.
Specifically, the standard ICN routing algorithm is not suitable for
opportunistic networking environments. As the network topology is very dynamic,
establishing FIB entries like in vanilla ICN routing makes little sense, as the
established next hops might not be meaningful after a while due to the mobility of nodes. Basically, what is missing is some additional information about the
appropriateness of any node to bring future Interest packets towards nodes
storing the requested content. In vanilla ICN this is implicit due to the stable
links between Internet routers. MobCCN collects statistics about encounters with nodes and data types to create a sort of
\emph{gradient-based information-centric graph} that will be subsequently exploited to forward
Interest packets and retrieve the desired content. Note that, as a overhead
reduction measure, we assume that data can be grouped into types, and that
Interest packets are generated for certain content types. Content types are a general concept, and
in the extreme case each data can be of its own type. Thus, MobCCN establishes routing
information to bring Interest packets towards known nodes that store data of
the requested type.

In MobCCN the gradient-based information-centric graph is based on the concept of
\emph{utility}, which represents the appropriateness of a node as a forwarder
for a specific type of content. Specifically, a given node $p$ that encounters other
nodes (including fixed sensors) providing data of type $i$, computes a
\emph{direct} utility value, which is a function of the frequency with which the mobile node ``encounters'' that type of data, say $ICT(p,i)$, i.e., function of the inter-contact time
between $p$ and any such node. Specifically, the direct utility of node $p$ for content
type $i$ is as follows:
\begin{equation}
	U_{p,i}^{(d)} = f\left(ICT(p,i)\right) = \frac{1}{ICT(p,i)} \; .
	\label{eq:direct-utility}
\end{equation}
Direct utilities are set to 0 for content types that are never encountered.
Utilities are advertised between nodes upon encountering. On the other hand, if node $p$ encounters a node $q$ that has not the content of type $i$, then node $p$ updates its \emph{indirect} utility for content type $i$ as a function of (i) the current utility of node $q$, and (ii) the inter-contact time with node $q$. Therefore, the indirect utility measures the frequency with which a given node can access a given type of data through a series of intermediate nodes, the last of which encounters directly IoT devices or mobile nodes providing that data type. Specifically, node $p$ computes an indirect utility value for content
$i$ through node $q$, as follows:
\begin{equation}
	U_{p,q,i}^{(ind)} = f\left(U_{q,i},ICT(p,q)\right) = \frac{1}{\frac{1}{U_{q,i}}+ICT(p,q)}\; , 
	\label{eq:indirect-utility}
\end{equation}
where $U_{q,i}$ is the utility advertised by node $q$ for content type $i$. Finally, the overall utility of node $p$ towards content type $i$ is updated as the
maximum over its direct utility and all indirect utilities through nodes it has encountered. In other words, it holds that
\begin{equation}
	U_{p,i} = \max_{j \in \mathcal{N}_p} \left\{U_{p,i}^{(d)},U_{p,j,i}^{(ind)}\right\} \; ,
	\label{eq:utility}
\end{equation}
where $\mathcal{N}_p$ is the set of nodes encountered by $p$. \footnote{Note that when a second sensing node is attached to the same vehicle/user, this represents another instance of the same application, which does not provide any new information with respect to the information provided by the first node. In other words, two nodes that are constantly in touch with each other will have ICT equal to zero, they will have exactly the same utility towards any data, and they will meet exactly the same set of other nodes. Therefore, they will essentially act as a unique node with respect to the MobCCN operations.}

\subsection{MobCCN routing}
\label{sub:mobccn-routing}
\noindent
MobCCN routing is responsible for creating the FIB entries for available content
types, and assigning utility values. This is accomplished through single-hop
Hello packets exchanged by nodes upon encountering.
For the sake of clarity, pseudocode~\ref{alg:hello_processing} lists conceptual steps of Hello message processing at each node. Without loss of generality, we assume that the pseudocode is executed by node $p$, during a contact with node $q$. 
Hello packets are used to advertise the node's utility for each specific content type it knows
about. Therefore, for content type $i$, any node $q$ advertises the value $U_{q,i}$
computed as in Equation~\ref{eq:utility}. When node $p$ receives such a packet
during contact with node $q$, it checks the FIB to see if an entry for content
type $i$ with ``face'' $q$ is already available (line 5). If not, it creates a new one (lines 6-8).
Then, it computes either the direct or indirect utility for content type $i$ through node $q$ (depending on whether content type $i$ is availabe at $q$ or not),
with Equations~\ref{eq:direct-utility} or \ref{eq:indirect-utility}, respectively, and stores it in the FIB entry
(updating the previous value, if the case -- lines 9-14). In addition, node $p$ updates its own utility towards content type $i$ using Equation~\ref{eq:utility}, which is the value it will advertise in the following Hello packets (line 15). Moreover, node $p$ also stores the
utility received by $q$ in an additional data structure \emph{Current Neighbours Utilities} (CNU) (lines 4, 18-19), that is used in the forwarding process explained next. This entry is removed when the contact with node $q$ finishes. 
\begin{algorithm}
		\begin{algorithmic}[1] 
			\Require $N_p$ \Comment{Set of nodes encountered by $p$}
			\Statex
			\State $\mathcal{M} \gets Hello(q).\mathtt{ContentList}()$ \Comment{list all content records in the Hello message from node $q$}			
			\While {$\mathcal{M}\neq\emptyset$}
				\State $i \gets \mathcal{M}.\mathtt{First}()$
				\State $\mathrm{CNU}.\mathtt{CreateEntry}(i,q)$ \Comment{Store utility of $q$ for content type $i$ in CNU}
				\State $\mathrm{FIB}.entry \gets \mathrm{FIB}.\mathtt{Find}(i)$
				\If {($\mathrm{FIB}.entry  = \emptyset )$)} \Comment{No routing information for content type $i$}
					\State $\mathrm{FIB}.\mathtt{CreateEntry}(i,q)$
				\EndIf
				\If { ($i$ in CS of node $q$)}
					\State $U_{p,i}^{(d)} = \frac{1}{ICT(p,i)}$
				\Else
					\State $U_{p,q,i}^{(ind)}  = \frac{1}{\frac{1}{U_{q,i}}+ICT(p,q)}$
				\EndIf
				\State $\mathrm{FIB}.\mathtt{UpdateEntry}(i,q)$ \Comment{Update FIB entry for content type $i$ through node $q$}
				\State $U_{p,i} = \max_{j \in \mathcal{N}_p} \left\{U_{p,i}^{(d)},U_{p,j,i}^{(ind)}\right\} $ \Comment{Overall utility of node $p$ towards content type $i$}
				\State $\mathcal{M} \gets \mathcal{M} -\{i\}$
			\EndWhile
			\State	\textbf{Upon} expiration of contact with $q$ \Comment{Remove the entry related to $q$ from CNU}
			\State $\quad\ \ \mathrm{CNU}.\mathtt{RemoveEntry}(i,q)$
		\end{algorithmic}
\caption{Processing of Hello messages received by node $p$}\label{alg:hello_processing}
\end{algorithm}

\subsection{MobCCN forwarding}
\label{sub:mobccn-forwarding}
\noindent
The forwarding actions are still the same used by vanilla CCN in
Figure~\ref{fig:CCN-forwarding}, with a simple algorithmic modification.
For the sake of clarity, pseudocode~\ref{alg:interest_processing} lists conceptual steps of Interest message processing at each node.

When node $p$ receives an Interest packet, it first checks whether the corresponding Data packet is stored locally in the Content Store. If this is the case, the Data is sent back to the requester (lines 2-4). Then, node $p$ generates a PIT entry (if there is
none for that content type) adding a ``face'' corresponding to the node that
sent the packet to it. If a PIT entry for the content type already exists, node $q$ is simply added as a new ``face'', and the algorithm completes (lines 5-8).
This is the way breadcrumbs are set. If the Interest has
to be forwarded (note that, as in vanilla ICN, this is the case only when a new PIT entry is created), node $p$ identifies the node with the highest utility it knows
about (lines 9-16). Remember that FIB values are computed using
Equation~\ref{eq:indirect-utility}, and thus, for any node $q$, they take into account the utility of node $q$, and
the inter-contact time between $p$ and $q$. This is correct if node $p$ and $q$
are not currently in contact. But if they are, the utility that node $p$ should
consider is the one that node $q$ advertises in its Hello packets ($U_{q,i}$).
Therefore, node $p$ behaves as follows. It first looks in CNU to the utilities
$U_{j,i}$ for all nodes $j$ currently in contact, and extracts the node with
the highest utility, if any. Then it performs the same operation looking at the FIB entries.
The node to be used as forwarder is the one having the highest utility among those obtained
from CNU and FIB. If this is among the nodes currently in contact, the Interest packet is forwarded (lines 11-12). Otherwise, the Interest packet is stored waiting for the next contact, and eventually forwarded (lines 13-15). Finally, if no FIB entry exists, the Interest packet is dropped, as in vanilla ICN (line 16-18).

Note that this algorithm can be made entirely compliant with vanilla CCN
forwarding. It is sufficient that, during contact with node $q$, node $p$
stores, in the FIB entry for content $i$ through node $q$, the value $U_{q,i}$
received from $q$ instead of the value $U_{p,q,i}^{(ind)}$ computed via
Equation~\ref{eq:indirect-utility}, which is instead used when node $p$ and $q$
are not in contact.

\begin{algorithm}
		\begin{algorithmic}[1] 
			\Require $E_p$ \Comment{Set of current neighbours of node $p$}
			\Statex
			\State $i \gets Interest.\mathtt{Name}()$
			\If {$\mathrm{Data} \gets \mathrm{CS}.\mathtt{Find}(i)$}
				\State \Return Data 
			\EndIf
			\If {$\mathrm{PIT}.\mathtt{Find}(i)$}
				\State $\mathrm{PIT}.\mathtt{AddFace}(i,q)$
		    \Else
		    	\State $\mathrm{PIT}.\mathtt{CreateFace}(i,q)$
				\If {$\mathrm{FIB}.\mathtt{Find}(i)$}
					\State find the node $j$ with the highest utility for content type $i$ in FIB and CNU tables
					\If {$j \in E_p$}
						\State forward Interest message to node $j$
					\Else
						\State store Interest message waiting for the next contact  
				\EndIf
				\Else
					\State drop  Interest message
				\EndIf
			\EndIf
		\end{algorithmic}
\caption{Processing of Interest messages received by node $p$}\label{alg:interest_processing}
\end{algorithm}

\subsection{MobCCN retransmission}
\label{sec:mobccn-retransmission}
Among the various mechanisms inherited by CCN, MobCCN also maintains the light retransmission mechanism of Interest packets used by the native CCN to cope with possible packet damage/loss and/or nodes carrying the Interest packets that disappear from the network, which works as follows. If multiple Interest packets for the same content arrive at one node, the node forwards the first Interest packet arrived, while it stores the "faces" from which the subsequent requests come in the corresponding PIT entry,  as explained in Section \ref{sub:mobccn-forwarding}.
Whenever the number of Interest packets for the same content exceeds a certain threshold and the requested content has not yet been received by the node, the node retransmits the first Interest packet received. This guarantees a reasonable level of protocol reliability with a low amount of additional network traffic. Note that the described mechanism can be enabled/disabled and that the retransmission threshold can be adjusted depending on the network environment. What we expect with the retransmission mechanism is an overall increase of the overall costs, but with a marginal improvement in terms of delivery rate and end-to-end delay. In the rest of the paper we refer to ``\emph{MobCCN}'' when the retransmission mechanism is active, while to ``\emph{MobCCN/noReTrans}'' when such mechanism is disabled.

\section{Performance evaluation settings}
\label{sec:peva-settings}

In the next sections of the paper, we provide a performance characterisation of MobCCN from a number of complementary standpoints. Therefore, in the rest of this section, we explain the settings of our performance evaluation study. Specifically, in Section \ref{sub:investigated-scenarios} we focus on the scenarios investigated in our simulation study and we present the performance indices used for comparison. In Section \ref{sub:ccn-protocols}, we present the protocols compared against MobCCN.  We then present the mobility model we have used (see Section\ref{sub:mobility}). Finally, in Section \ref{sub:content-requests} we describe the content request patterns in our simulations. 

The performance evaluation is carried out through a software prototype of our MobCNN protocol and of all the alternative protocols described hereafter, as an extension of
CCN-lite~\cite{ccn-lite}, an open source lightweight implementation of the CCN protocol with
compact codebase and reduced memory footprint. These features make CNN-lite
particularly suitable for IoT environments and resource-constrained devices, in
general. Another advantage of CCN-lite is its ability to support multiple
hardware/software platforms. Specifically, the same implementation can be easily
executed in various operating systems (Linux, Android OS), embedded systems
(Arduino), and simulation environments (OMNeT++). In the following section we
show results obtained using the OMNeT++ tool, while experiments in a real-life
scenarios are left as future work.

\subsection{Simulation scenarios and performance indices}
\label{sub:investigated-scenarios}

In our simulation study, we focus on two complementary configurations (see Table ~\ref{tab:settings}).
On the one end, we investigate a small-scale network composed of 10 nodes moving in a square area of side 1000m. Among them, four nodes hold the available content and a variable number of nodes request the content. Hereafter we refer to this network configuration to as \emph{Scenario A}. In this scenario we study the protocol performance in terms of network resources consumed. To this aim, we vary the traffic load in two directions, i.e., both increasing the number of nodes requesting content and the number of requests per node. As a result, by neglecting issues related to network congestion (as it is typical in the opportunistic networking literature), we isolate the effect on delivery and network overhead performance of the MobCCN algorithms in comparison with reference Epidemic-style CCN protocols, and such comparison is relevant also in such a small-scale configuration.

We then investigate a larger network composed of 30 nodes moving in the same square area. Here, we assume a fixed number of nodes holding the available content and a fixed number of nodes request the content. We refer to it as \emph{Scenario B}. The objective of this scenario is to evaluate the protocol performance with a different node density and mobility patterns followed by nodes (see Section \ref{sub:mobility}), and to study additional performance metrics that in a small context would remain hidden, such as number of hops taken by Data packets. We complement the network resource analysis with the study of device resource usage to see how many device resources MobCCN is able to save.

\begin{table}[h!]
  \centering
  \caption{Simulation settings.}
  \label{tab:settings}
  \begin{tabular}{|l|c|c|}
    \hline 
    \rowcolor[gray]{.8}\multicolumn{3}{|c|}{\textbf{Mobility parameters}}\\
     \hline  \hline
     & \cellcolor[gray]{0.85}Scenario A & \cellcolor[gray]{0.85}Scenario B \\
     \hline
    Area &     \multicolumn{2}{|c|}{1$km\times$1$km$ }\\
    \hline 
    mobility model & \multicolumn{2}{|c|}{HCCM~\cite{Boldrini20101056} }\\
    \hline 
    n. mobile nodes & 10 & 30 \\
    \hline
    n. mobile communities & 1 & 3 \\
    \hline
    n. travellers & 0 & 3 \\
    \hline
   communication range & 20$m$ &5$m$\\
    \hline
    avg. speed & \multicolumn{2}{|c|}{$\mathcal{U}(1,1.86)~m/s$} \\
    \hline  \hline
    \rowcolor[gray]{.7}\multicolumn{3}{|c|}{\textbf{Traffic parameters}}\\
    \hline  \hline
         &  \cellcolor[gray]{0.85}Scenario A &  \cellcolor[gray]{.85}Scenario B \\
     \hline
    n. producers & 4 & 3\\
    \hline
    n. consumers & 1,2,5 & 3\\
    \hline
    n. content types & 10 & 4\\
    \hline
    n. chunks per content type & 25 & 5\\
    \hline
    n. requests per consumers & 50,100 & 40\\
    \hline
    distribution req. times  & geometric ($p=2.7$ hours)  & exponential ($\lambda=16.6$ minutes) \\
    \hline
    simulation time  &  \multicolumn{2}{|c|}{24~hours } \\
    \hline
    req. start time  &  \multicolumn{2}{|c|}{12~hours } \\
    \hline
    req. end time  & \multicolumn{2}{|c|}{ 22~hours } \\
    \hline
     \end{tabular}
\end{table}

In all the performed simulations we assume that nodes have infinite bandwidth during contacts (or, that the available bandwidth is much larger than what required to transfer any set of messages between any two nodes), and that the local caches used to store messages at intermediates nodes is of unlimited size. Note that, this is a typical assumption in the analysis of opportunistic networking protocols, and it is known to favour resource-hungry protocols like Epidemic.

We use several performance indices for the simulation comparison. In both scenarios, we analyse the performance in terms of (i) delivery rate, (ii) end-to-end delay, and (iii) traffic overhead. Specifically, the delivery rate is the percentage of Data packets received by requesting nodes. As we assume that no packet is lost in the network, the only reason why a Data packet may not reach the requesting node is that the it, or one of its corresponding Interest packets, have not yet been delivered to their final destinations at the end of the simulation. We took care of avoiding corner cases due to content being request close to the end of simulation runs, as explained in Section~\ref{sub:mobility}.  Then, the end-to-end delay represents the time interval between the generation of the Interest packet by the consumer and the reception of the corresponding Data packet. This index is computed only for those Data packets successfully received.  
As far as the traffic overhead, in Scenario A, we simply measure the amount of traffic (in bytes) generated by all nodes in the network, for sending Interest, Data and control traffic required by the specific protocol under consideration. Conversely in Scenario B, we isolate the traffic sent due to Interest and Data packet forwarding from the control traffic. By doing this we have a clear picture of the amount of requests and data packets circulate, but at the same time we provide a measure of how much bandwidth MobCCN consumes to exchange information upon which to create the gradient-based information-centric graph. 
Moreover, in the Scenario B we consider three additional performance indices: (i) the number of hops taken by Data packets; (ii) the number of copies of the Data packets that reach the requesting nodes, and (iii) the percentage increase of cache utilization with respect to its initial size. Index (i) allows us to compare protocols in terms of the number of nodes involved in the forwarding process (and thus, the number of nodes that are requested to ``consume'' resources). Index (ii) measures the number of useless copies that requesting nodes are receiving. Finally, index (iii) provides a measure of how many local storage resources are consumed at each node, and it is computed as: 

\begin{equation} 
 C_{u} = \frac{C_{f}-C_{i}}{C_{i}} * 100   
\label{eq:cache-utilization} 
\end{equation}  

where $C_{i}$ represents the total number of contents stored at all the nodes at the beginning of the simulation, and $C_{f}$ represents the same quantity at the end of the simulation. 

We also run simulations with and without caching of Data packets. The effect of caching is to increase the number of nodes that can serve Interest packets, and thus evaluating the protocols with and without caching allows us to show to what extent having more copies of requested data modifies the performance, both in terms of end-to-end delay and in terms of network overhead. 

For each simulation we perform multiple runs and we present the average value of the statistic. Confidence intervals are computed with 95\% confidence level.

\subsection{Alternative protocols}
\label{sub:ccn-protocols}

In this section we describe the benchmarks used in the performance evaluation section. Specifically, we compare MobCCN against different protocols all based on modifications of the original Epidemic forwarding protocol for opportunistic networks~\cite{vahdat2000epidemic}. This choice is driven by the fact that a significant part of the literature about ICN for IoT (see Section~\ref{sec:related}) claims that generating routing traffic to populate FIBs (as in the original CCN definition and in MobCCN) is not worth in mobile environments, as the environment is too dynamic, and therefore maintaining FIB entries is too costly or even meaningless. Instead, Epidemic schemes are used to spread Interest packets. As MobCCN follows the opposite approach, testing it against different protocols based on Epidemic routing allows us to understand whether this is a sensible choice.

\subsubsection{Ideal Epidemic}
\label{sub:id-epidemic}
\emph{Ideal Epidemic} is a protocol where \emph{both} Interest packets and Data packets are flooded in the network according to a classic epidemic replication scheme. Therefore, in this version of Epidemic-like CCN routing, both the Interest packets and the corresponding Data packets reach the intended destinations along the optimal (quickest) path, and both achieve the minimum end-to-end delay. This is true under our working hypothesis of having infinite bandwidth during contacts and unlimited cache size at nodes. This comparison allows to evaluate the performance of MobCCN in terms of end-to-end delay, with respect to the lower bound achievable by opportunistic networking protocols. Indeed, it is well-known from the opportunistic networking literature~\cite{DBLP:journals/winet/SpyropoulosRTOV10} that, when network congestion is not taken into account, Epidemic~\cite{vahdat2000epidemic} is the protocol that guarantees the best end-to-end delay for a given source-destination pair. This is because, starting from the source, Epidemic essentially floods the network, and therefore it finds - among many others - the quickest path to the destination. Comparing opportunistic networking protocols against Epidemic in terms of end-to-end delay is thus a way to show how far a given protocol is from the optimal - ideal - end-to-end delay achievable in the network.

\subsubsection{Epidemic-CCN-1copy}
\label{sub:epidemic-CCN-1copy}
Using \emph{ideal Epidemic} we compare against one possible edge of the spectrum, i.e., we compare against a protocol that uses as many network resources as possible, to achieve the minimum possible end-to-end delay and maximal delivery rate. Then, we compare MobCCN against a protocol at the other edge of the spectrum, i.e., a protocol that uses minimal resources. To this end, we define a protocol, hereafter referred to as ``\emph{Epidemic-CCN-1copy}'' that behaves as follows. Interest packets are forwarded epidemically, but only one copy of the packet is allowed to exist in the network at any point in time. Therefore, when an intermediate node encounters another node, the Interest packet is forwarded to the latter with a probability $p$ (set in our experiments to 0.5). If forwarding occurs, the node deletes the Interest packet locally, and does not spread it epidemically any further\footnote{The forwarding of Interest packet is to some extent analogous to a random walk process. }. Moreover, Data packets are sent to requesters as per the original CCN mechanism, i.e., by using the breadcrumbs established by the Interest packet propagation. In this way, we obtain a ``single-copy'' protocol based on Epidemic, where only one copy of both Interest and Data packets are allowed in the network. Note that, also MobCCN is single copy in this sense. Epidemic-CCN-1copy does not generate any control traffic, and is the version based on Epidemic dissemination that drastically reduces the network traffic. Possibly, this may be paid with longer end-to-end delays and lower delivery ratios. The purpose of this part of the analysis is thus to analyse the effectiveness of its utility-based routing scheme of Interest packets, with respect to a protocol that does not use any concept of utility to find opportunistic paths between data and nodes requesting them.  

\subsubsection{Epidemic-CCN-1copy/noReTrans}
\label{sub:epidemic-noReTrans}
Finally, we consider a lighter version of Epidemic-CCN-1copy, hereafter referred to as ``\emph{Epidemic-CCN-1copy/noReTrans}'', where Interest packets for the same content are not retransmitted. 
Specifically, Epidemic-CCN-1copy/noReTrans behaves as follows. When an Interest packet arrives at an intermediate node, the Interest packet is forwarded  with probability $p$ if and only if no Interest packet for the same content has been forwarded earlier. Data packets are then routed back to the consumer as in the Epidemic-CCN-1copy (i.e., by using breadcrumbs established by the Interest packets). This mechanism guarantees that a unique copy of each request for the same content is allowed to circulate in the network at any time. This version of Epidemic generates the least possible network traffic with a possible pay-off in terms of increase of end-to-end delays and decrease of delivery ratios. What we expect from the use of  Epidemic-CCN-1copy/noReTrans is a further decrease of the network costs, but with a noticeable deterioration of delivery performance. Furthermore, the direct comparison between MobCCN and this light version of Epidemic allows to highlight once again the wide difference between utility-based routing protocols and Epidemic-like ones.

\subsection{Mobility model}
\label{sub:mobility}

To simulate opportunistic contacts between mobile nodes we use a trace-based
mobility model. Specifically, we use the Home-cell Community-based Mobility Model (HCMM) mobility model, a popular community-based human model~\cite{Boldrini20101056}, to generate plausible mobility patterns
for users' personal mobile devices. 
In Scenario A we set HCMM so that 10 nodes
move in square area of side 1000m. This corresponds to simulate movements of
10 nodes belonging to the same social community (e.g., colleagues at work),
which therefore move according to the same statistical process in the simulation
area. In this configuration, HCMM results in a mobility model equivalent to
the Random Waypoint Model (RWP), implemented in such a way to avoid its well-known
issues related to filtering the initial non-stationary phase. For movements
inside a given social group, this has been shown to be a sufficiently realistic
model~\cite{Boldrini20101056}. Nodes move at a speed randomly selected in the range $[1,1.86]$~m/s, and the
transmission range is 20m. Therefore, the resulting network is extremely sparse.

Scenario B aims to reproduce the mobility of nodes that spend most of the time with nodes of the same social community (e.g., colleagues at work) but that have sporadic contacts with nodes belonging to external social communities (e.g., for a coffee break). To this aim, we set HCMM so that 30 nodes move in the same square area with the same average speed as before, but uniformly divided into three different social communities. They move inside their home community, but three nodes among them (one per each community) perform as "travellers", i.e., have social relationships with the other two foreign communities.

\subsection{Content request patterns}
\label{sub:content-requests}

In our tests, we do not simulate the fixed
sensor nodes but we assume that at the beginning of the simulation a number of
mobile nodes have the content already in their caches. The direct utility of
this initial forwarders is randomly selected to a high value.

In Scenario A, we assume that four nodes in the network hold the available content, that 
10 content types are available, and each content type contains 25 chunks (see Table ~\ref{tab:settings}). Content types are
allocated according to a uniform distribution on the four ``content provider''
nodes at the beginning of the simulation. During the simulation, a variable
number of other nodes (hereafter, ``consumers'') ask for contents, requesting
a variable number of chunks per request. The content type and set of chunks is selected
according to a uniform distribution. As far as the request generation pattern,
we assume that the requests times are generated according to a geometric distribution (as we
consider discrete time) with mean equal to 10000 seconds (i.e., 2.7~hours). Specifically, we generate a number of request events
equal to the total number of requested chunks. As all requests are generated
within the same simulation time interval, the effect of requesting additional
chunks is to concentrate more the requests in time, around the average value of the geometric distribution. As explained before, the effect of both
increasing the number of consumers, and increasing the number of requests chunks,
is to test the performance of MobCCN for a variable traffic load. Finally, there might be a partial overlap of contents requested by consumers, as each consumer can request any chunk at random. This has the effect of making the overlapping contents more popular, and thus - in case of caching of Data packets - more diffused in the network.

In Scenario B, we assume that one node for each community is a content provider holding 4 content types, each containing 5 chunks, while one node for each community is a consumer that makes 40 requests (see Table ~\ref{tab:settings}). Content types are uniformly allocated on the three providers at the beginning of the simulation as before, while each request (i.e., a content-chunk pair) is selected according to a uniform distribution so that 50\% of the total requests made by each consumer is for content within its social community, while the remaining 50\% is uniformly distributed among the other two foreign communities. Consumers generate requests so that inter-request times follow an exponential distribution with $\lambda=1000sec$. This means that each consumer generates, on average, a request for a different content-chunk pair every 16.6 minutes.  Also in this scenario there might be a partial overlapping among the requests made by the different consumers.

Finally, simulations last for 24 hours in both scenarios. We used the first 12 hours as warm-up period, and the consumers start requesting chunks only after this warm-up period. Furthermore, we do not generate requests in the last 2 hours, to avoid drop rates due exclusively to truncation of simulation time. This makes sure that, when Data packets are not received, this is due to either Interest of Data packets being trapped in regions of the network from where they cannot reach the intended destinations, as a side effect of the nodes' mobility patterns, i.e., this allows us to measure packet drops due to wrong routing decisions taken by the analysed protocols.

\section{Performance evaluation results}
\label{sec:results}
In this section we evaluate the performance of the MobCCN protocol described in Section \ref{sec:mobccn}.  First, we focus on the small-scale scenario where we mainly analyse the system performance with different network traffic loads, then we move to the larger network scenario where we confirm the performance advantages of the use of MobCCN previously shown and we highlight its node resource saving in terms of caching space and duplicate contents.

\subsection{Scenario A}
\label{sub:comparison-scenarioA}
Scenario A refers to a small-scale network composed of 10 nodes, where a variable number of consumers request the available contents held by the four producers. First, we compare MobCCN against ideal Epidemic to see the MobCCN behaviour with respect to a protocol that guarantees the highest performance indices at the cost of a high network resource consumption. Then, we compare MobCCN with Epidemic-CCN-1copy, which, on the contrary, uses the minimal resources. In both cases we use (i) delivery rate, (ii) end-to-end delay, and (iii) network overhead as performance metrics.

\subsubsection{Comparison against ideal Epidemic}
\label{sub:mobcnn-vs-ideal-epidemic}

\begin{figure}[htpb]
	\centering
	\includegraphics[trim={0, 0, 0, 0}, clip, scale=0.3]{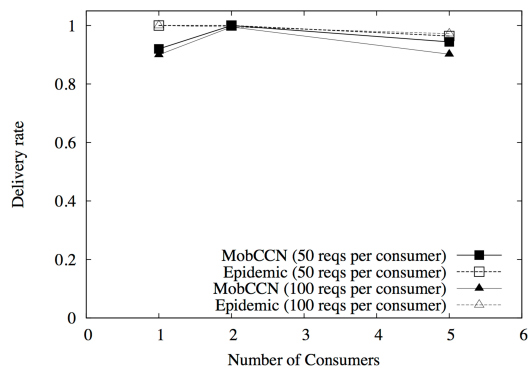}
	\caption{Delivery rate}
	\label{fig:deliveryRate_scen1}
\end{figure}

Figure~\ref{fig:deliveryRate_scen1}
compares the delivery rate for a varying number of consumers (nodes requesting
the content). We can clearly see that the performance of MobCCN and Epidemic are
basically equivalent. Both protocols achieve almost 100\% delivery rate in all
configurations. This was expected for Epidemic, given the configuration of our
simulations. For MobCCN, this confirms that the routing and forwarding
algorithms are able to identify suitable multi-hop opportunistic paths, such
that almost all contents can be delivered within the simulation time.

\begin{figure}[tpb]
	\centering
	\includegraphics[trim={0, 0, 0, 0}, clip, scale=0.3]{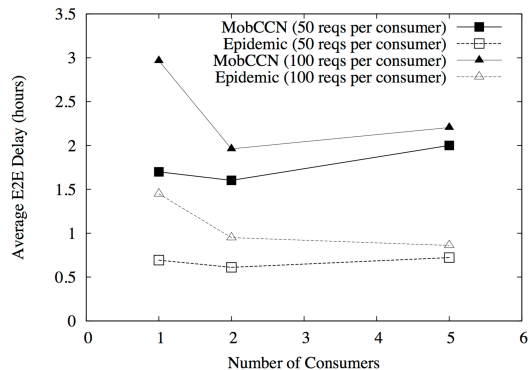}
	\caption{End-to-end delay}
	\label{fig:e2e_scen1}
\end{figure}

Figure~\ref{fig:e2e_scen1} shows the end-to-end delay. In this case, clearly
Epidemic outperforms MobCCN. This was also expected. Specifically, as we
neglect all network congestion issues, Epidemic is able to explore all possible
paths to propagate Interests, and to bring contents back to requesting nodes, and thus it finds the optimal ones.
MobCCN explores a single path (based on its routing and forwarding algorithm),
that may be sub-optimal, as it is computed based on average statistics on
inter-contact times. However, the delay is about three times that of
Epidemic, which in this configuration can be considered as a good result. Note
that in more realistic conditions, when network congestion is considered,
the gap is expected to be significantly lower, as the network would be much more
congested for Epidemic. Finally, note that the increase of end-to-end delays for
the case of one consumer when requesting more chunks is due to the fact that
additional chunk requests concentrate in a time window while the consumer is in
a particularly sparse area of the network, and thus needs more time to receive
the requested chunks.

\begin{figure}[hpb]
	\centering
	\includegraphics[trim={0, 0, 0, 0}, clip, scale=0.3]{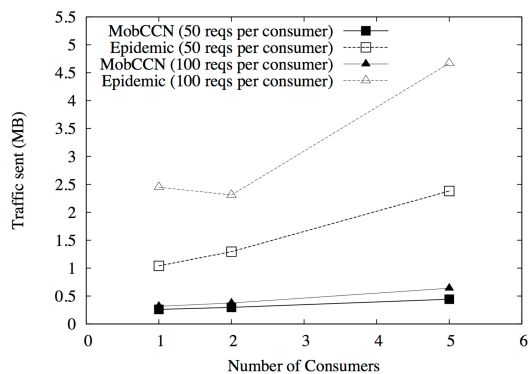}
	\caption{Total traffic sent}
	\label{fig:trafficSent_scen1}
\end{figure}

Finally, Figure~\ref{fig:trafficSent_scen1} compares the total traffic (including Data and Interest packets) that is transmitted over the network. The curves
clearly show the great advantage of MobCCN from this stanpoint. Specifically, overhead increases
linearly both for MobCCN and Epidemic, due to (i) the higher number of Interest
packets and (ii) the higher number of Data packets generated when both the
number of customers and the number of requested chunks increases. However, the
performance difference is striking, as MobCCN cuts the overhead by one order of
magnitude. This is a very important point for evaluating MobCCN. As anticipated,
thanks to the use of opportunistic routing policies, MobCCN generates a limited
amount of traffic to populate FIB tables, but this pays off very well, as it
does not need to generate excessive traffic to spread epidemically Interest
packets. Moreover, Data packets are routed back along one unique route, and this
avoids useless replication of Data packets along multiple routes as in the case
of Epidemic.

\subsubsection{Comparison with Epidemic-CCN-1copy}
\label{sub:comparison-epidemic-ccn}
\begin{figure}[tb] 
\centering
    \subfloat[50 requests \label{fig:deliveryRate_scen2_50}]{%
      \includegraphics[trim={0cm 0cm 0cm 0cm},clip,angle=-90,width=0.5\textwidth]{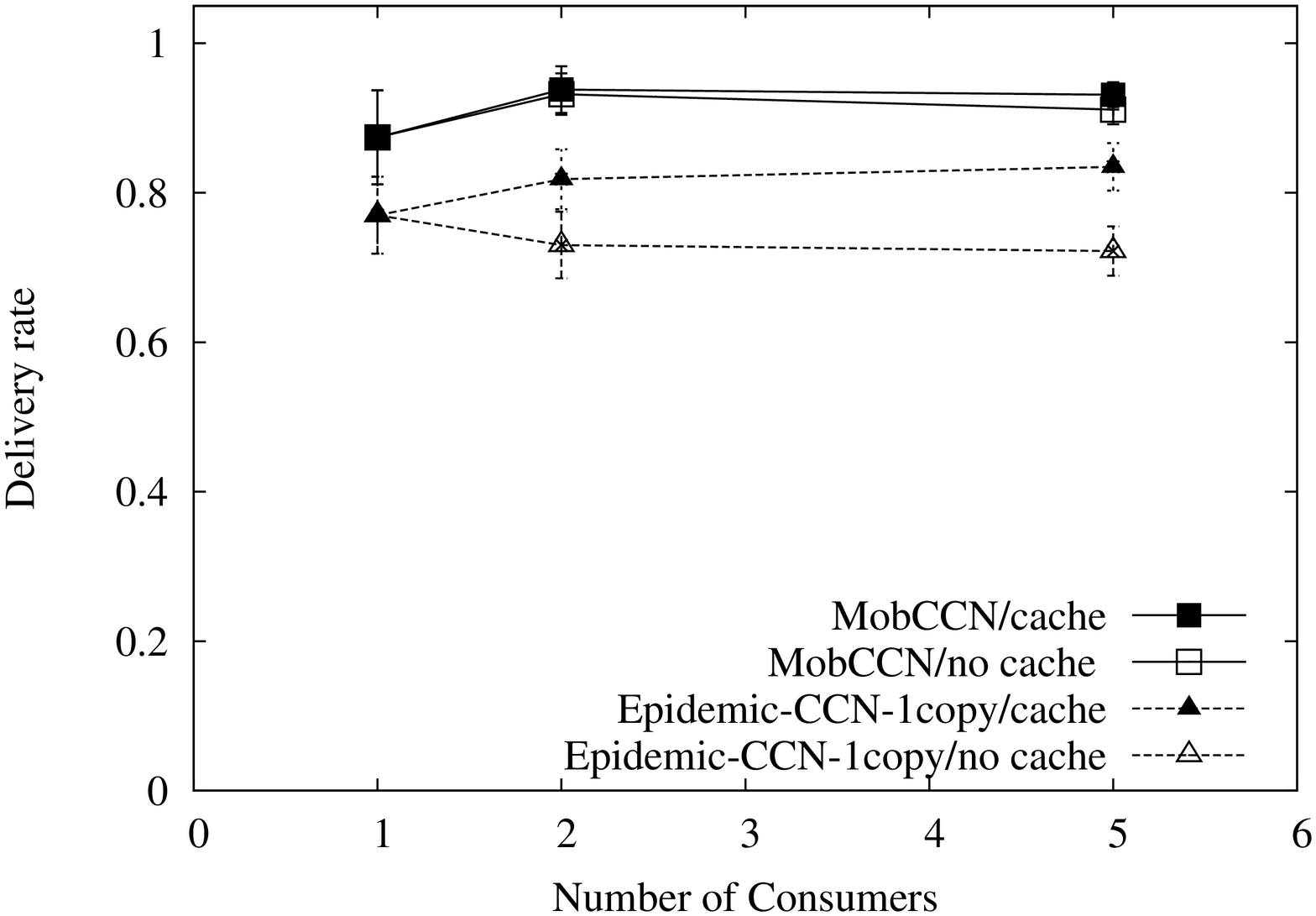}
    }
%
%
    \subfloat[100 requests\label{fig:deliveryRate_scen2_100}]{%
      \includegraphics[trim={0cm 0cm 0cm 0cm},clip,angle=-90,width=0.5\textwidth]{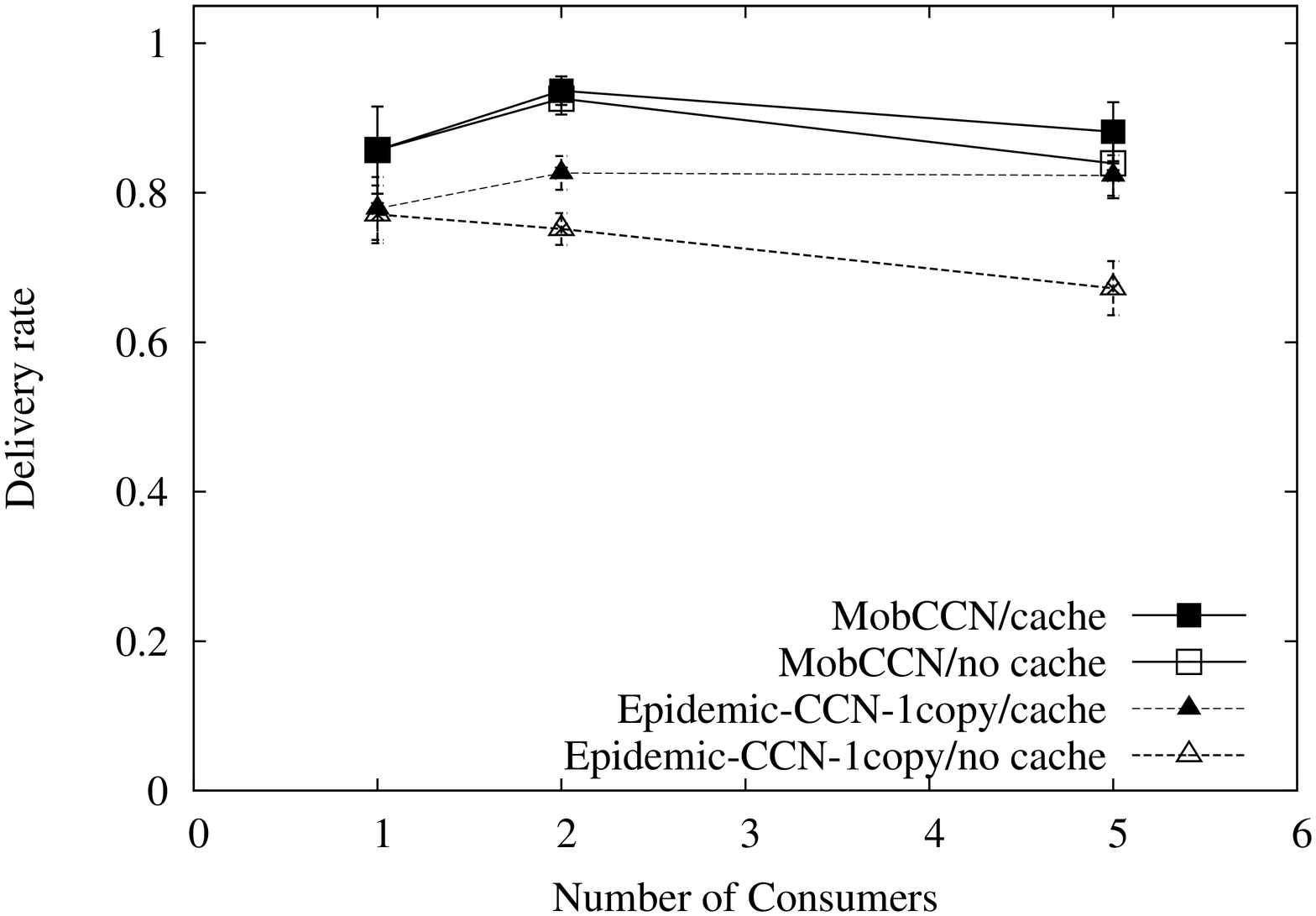}
      }
\caption{Delivery rate.}
\label{fig:deliveryRate_scen2}
\end{figure}

Figure~\ref{fig:deliveryRate_scen2} shows the delivery rate of MobCCN and Epidemic-CCN-1copy for varying number of consumers (nodes requesting the content) and different loads (total number of requests per node). In the same figure we also asses the impact on the delivery rate of the availability of a local cache that is used to store the Data packets that are forwarded by the node. Differently from the results shown in Figure~\ref{fig:deliveryRate_scen1}, we can observe that MobCCN significantly outperforms Epidemic-CCN-1copy, especially when operating without a data cache. More precisely, MobCCN closely approaches a 100\% delivery rate in the case of 50 content requests per node and it is above 90\% delivery rate in the case of 100 content requests per node. On the contrary, Epidemic-CCN-1copy with no cache obtains a delivery rate as low as 60\% when there are 5 consumers that generate 100 content requests each. Note also the slight increase of the average delivery rate when the number of consumers grows from one to two. However, the results are within their respective confidence intervals. Thus, they are statistically equivalent.  

Furthermore, the performance of MobCNN with and without a local cache are practically equivalent. This is due to the fact that the MobCCN routing is able to populate FIB tables with the most suitable next hop to reach a specific content based on observed contact patterns, while vanilla ICN routing is only able to follow the reverse path of the first Interest packet that reached a node producing (or having a copy of) that content. It is also important to point out that the performance of Epidemic-CCN-1copy degrades up to 20\% in the considered scenario if the nodes are not using a data cache. This can be explained by observing that in this case only the producers hold a copy of the content and Epidemic-CCN-1copy routing relies on a random search of the content providers as Interest packets are delivered to newly discovered contacts at random. Finally, Epidemic-CCN-1copy obtains a delivery rate that is lower than ideal Epidemic (see Figure~\ref{fig:deliveryRate_scen1}). This is due to the fact that Epidemic-CCN-1copy tries to limit the protocol overhead by not replicating the same Interest packet to all encountered nodes as each node forwards the received Interest packet to only one of its current neighbours.\footnote{Note that, even if the generation of requests ends with a delta from the end of the simulation greater than the average end-to-end delay, the protocol performance is essentially identical. Specifically, the delivery rate increases to a maximum of 3\% for Epidemic and 4\% for MobCCN. Importantly, the ranking among the compared protocols is not affected.}

\begin{figure}[] 
\centering
    \subfloat[50 requests \label{fig:e2e_scen2_50}]{%
      \includegraphics[trim={0cm 0cm 0cm 0cm},clip,angle=-90,width=0.5\textwidth]{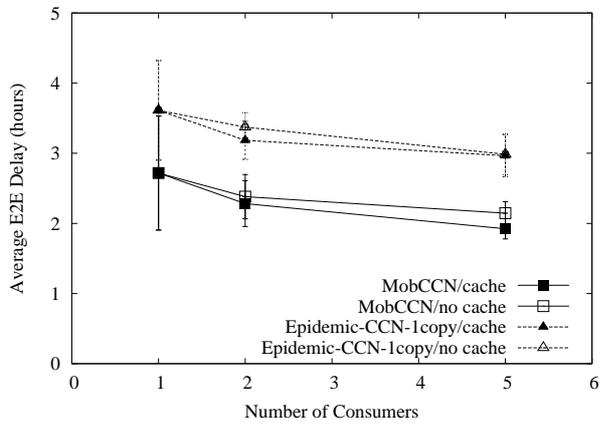}
    }
%
%
    \subfloat[100 requests\label{fig:e2e_scen2_100}]{%
      \includegraphics[trim={0cm 0cm 0cm 0cm},clip,angle=-90,width=0.5\textwidth]{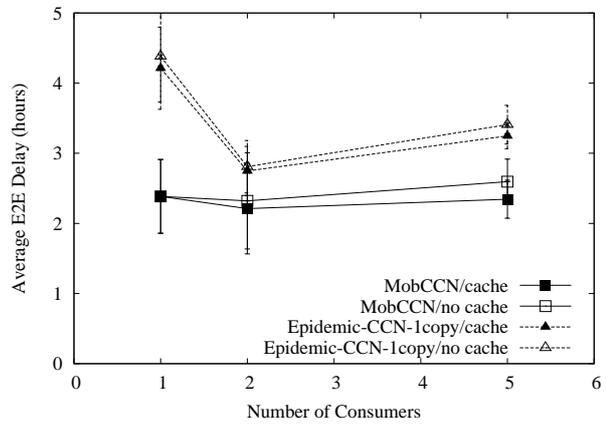}
      }
\caption{End-to-end delay.}
\label{fig:e2e_scen2}
\end{figure}
\begin{figure}[] 
\centering
    \subfloat[50 requests \label{fig:trafficSent_scen2_50}]{%
      \includegraphics[trim={0cm 0cm 0cm 0cm},clip,angle=-90,width=0.5\textwidth]{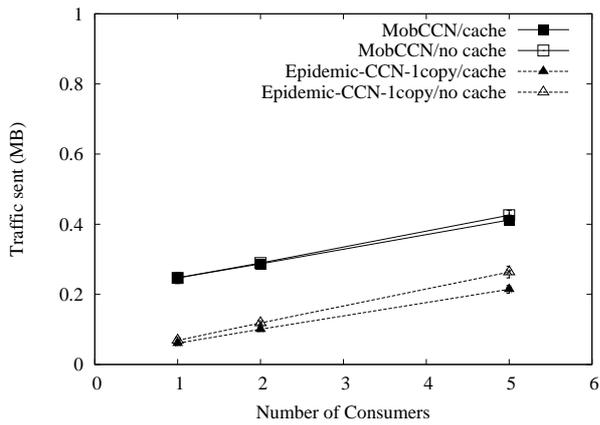}
    }
%
%
    \subfloat[100 requests\label{fig:trafficSent_scen2_100}]{%
      \includegraphics[trim={0cm 0cm 0cm 0cm},clip,angle=-90,width=0.5\textwidth]{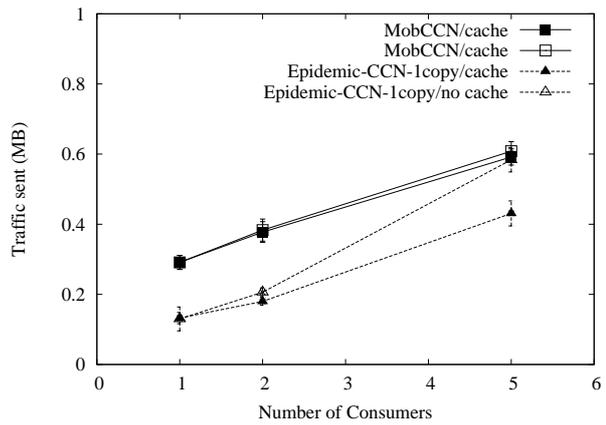}
      }
\caption{Total traffic.}
\label{fig:trafficSent_scen2}
\end{figure}
Figure~\ref{fig:e2e_scen2} compares the end-to-end delays of MobCCN and Epidemic-CCN-1copy in the same scenarios that are shown in Figure~\ref{fig:deliveryRate_scen2}. Also with respect to this performance index, MobCCN outperforms Epidemic-CCN-1copy. Epidemic-CCN-1copy tries to limit the protocol overheads with respect to ideal Epidemic by limiting the number of message replications. This reduces the efficiency of the Epidemic routing, which not only provides a lower delivery rate than MobCCN but it also finds sub-optimal paths thus resulting in longer end-to-end delays. It is specifically interesting to point out that this result was not necessarily expected in our configuration. Specifically, HCMM with one social community (which is the configuration we used in the paper) is known~\cite{Boldrini20101056} to be statistically equivalent to the Random Waypoint model, and therefore all nodes encounter according to the same statistical process. This might suggest that a random choice of next hops should be very efficient. However, MobCCN is able to consistently exploit the intrinsic variability of the mobility patterns, i.e., it is able to discriminate between different possible forwarders when differences in the mobility patterns exist due to the specific realisations of the underlying stochastic process.

Finally, Figure~\ref{fig:trafficSent_scen2} compares the total traffic sent by MobCCN and Epidemic-CCN-1copy. As expected, in this case Epidemic-CCN-1copy generates less traffic than MobCCN not only because the delivery rate of Epidemic-CCN-1copy is lower but also because it adopts a restrictive policy to reduce the number of forwarded messages. We remind that, differently from conventional Epidemic, Epidemic-CCN-1copy uses the CCN routing to route back Data packets with the requested data. Thus, not only Interest packets but also Data packets are not duplicated. As explained previously, MobCCN introduces control routing messages to distribute utility information and populate FIB tables. However, this additional protocol overhead is limited and it is balanced by the more efficient forwarding of Interest packets that is ensured by use of opportunistic routing policies.

\subsection{Scenario B}
\label{sub:comparison-scenarioB}

As explained in Section \ref{sub:investigated-scenarios},  Scenario B refers to more complex network composed of 30 nodes divided in three communities (i.e., 10 nodes/community), each with one producer, one consumer and one traveller. 
In this scenario we evaluate the performance of MobCCN with respect to all the three benchmarks presented in Section \ref{sub:ccn-protocols}, i.e., ideal Epidemic, Epidemic-CCN-1copy, and Epidemic-CCN-1copy/noReTrans. In addition, we include MobCCN/noReTrans (i.e., where the native CCN retransmission mechanism of old received Interest packets for the same content is disabled) to understand how the native CCN retransmission mechanism affects the protocol performance. First, we compare the protocols in terms of delivery performance (see Section \ref{sub:delivery-performance-scenarioB}), then we focus on the resource consumption (see Section \ref{sub:resource-consumption-scenarioB}).

\begin{figure}[t]
	\centering
	\includegraphics[trim={0cm 0cm 0cm 0cm}, clip, angle=-90,scale=0.3]{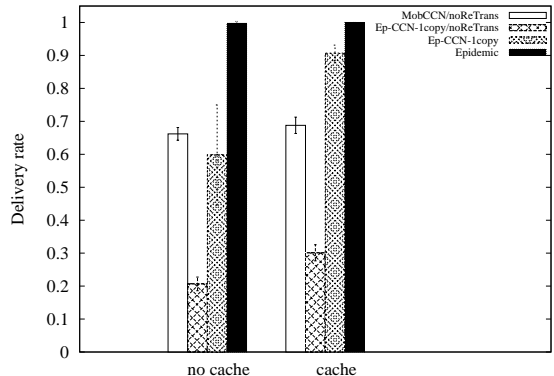}
	\caption{Delivery rate}
	\label{fig:deliveryRate_scen3comm}
\end{figure}

\subsubsection{Delivery performance}
\label{sub:delivery-performance-scenarioB}

As first analysis, we evaluate the performance of the two versions of MobCCN (i.e., with and without the retransmission mechanism) in terms of delivery rate and end-to-end delays. The behaviour of MobCCN without the retransmission mechanism (i.e., MobCCN/noReTrans) is similar of those of MobCCN. Specifically, we observe that the delivery rate increases of in a range of [0.03-0.04] when retransmission is enabled, independently of the use of data caching. As far as the end-to-end delays, again there is a small reduction when the retransmission is enabled, but in practice they are equal. We can conclude that, from a delivery performance standpoint, the use of retransmission mechanism does not produce any significant advantages, while it contributes to increase the network overhead, as we show in Section \ref{sub:resource-consumption-scenarioB}. For this reason, in the rest of the paper we refer to MobCCN/noReTrans results only.

\begin{figure}[t] 
\centering
    \subfloat[\label{fig:deliveryRate_LocalForReq_scen3comm:deliveryRate_LocalReq}]{%
      \includegraphics[angle=-90,width=0.5\textwidth]{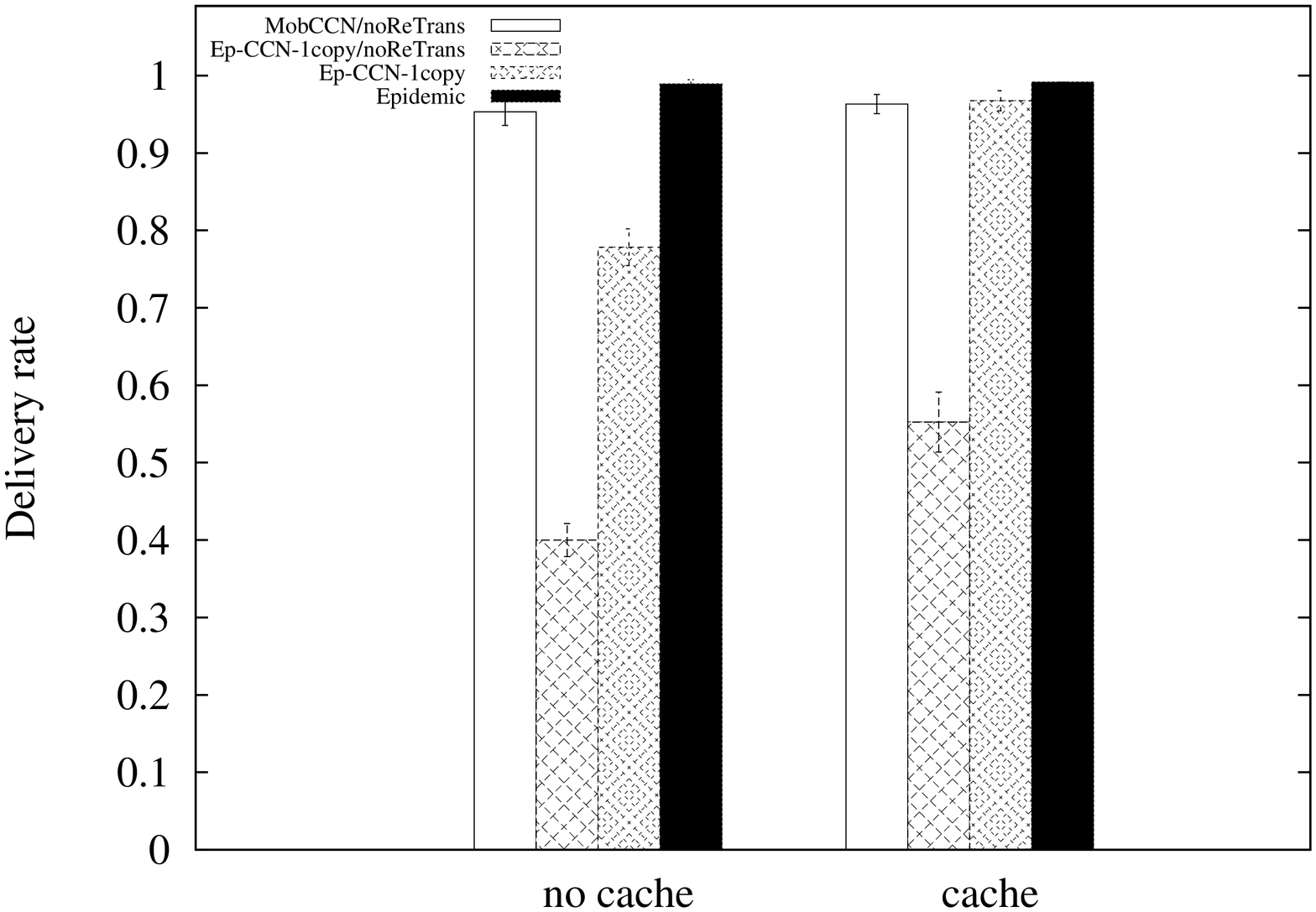}
    }
%
%
    \subfloat[\label{fig:deliveryRate_LocalForReq_scen3comm:deliveryRate_ForReq}]{%
      \includegraphics[angle=-90,width=0.5\textwidth]{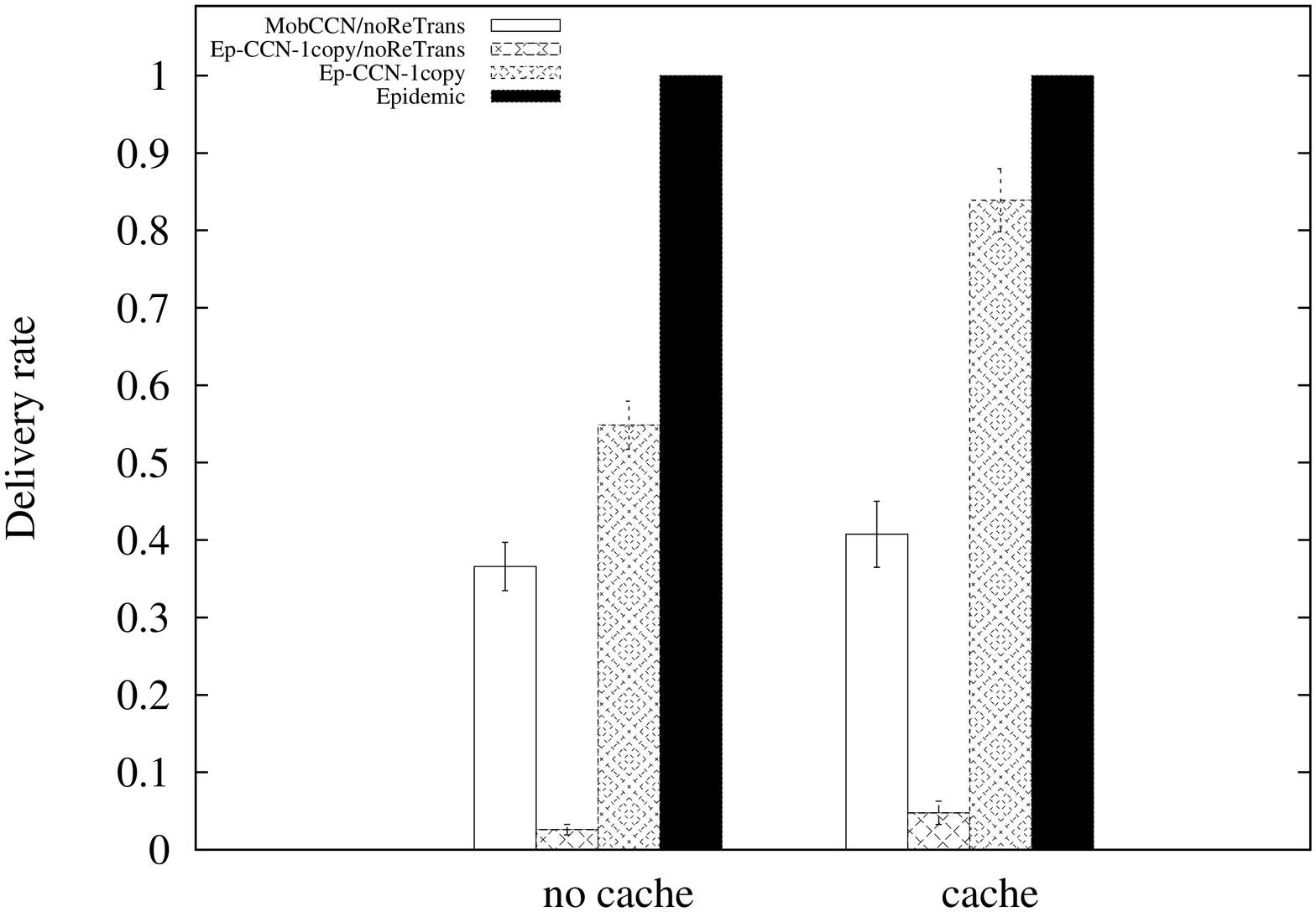}
      }
\caption{Breakdown of the delivery rate: home community (\ref{fig:deliveryRate_LocalForReq_scen3comm:deliveryRate_LocalReq}), and foreign communities. (\ref{fig:deliveryRate_LocalForReq_scen3comm:deliveryRate_ForReq}).}
\label{fig:deliveryRate_LocalForReq_scen3comm}
\end{figure}

Figure \ref{fig:deliveryRate_scen3comm} compares the delivery rate of the four protocols. Ideal Epidemic achieves the highest performance. This result is expected as Interest and Data packets are replicated on all encountered nodes, hence paths are optimal. However, when the number of message replications is limited, MobCCN/noReTrans outperforms the other two versions of Epidemic. This result is important as it confirms that MobCCN performs well delivering a large part of requested contents even when a single copy of Interest packet per different content circulates in network. When the caching mechanism is running, MobCCN/noReTrans still guarantees a delivery rate of about 0.7, outperforming the corresponding Epidemic version (i.e., Epidemic-CCN-1copy/noReTrans). However,  the performance of Epidemic-CCN-1copy and Epidemic is higher of the performance of MobCCN/noReTrans, but this comes at the cost of higher resource consumption, as it will be explained later. Note also the general increase of the protocol performance when Data packets are cached by nodes. Finally, another interesting result is that MobCCN/noReTrans without caching achieves almost the same delivery rate of MobCCN/noReTrans with cache. This is another advantage of MobCCN/noReTrans without cache, because to achieve similar or better performance a more expensive Epidemic-like protocol must be used (that is, at least Epidemic-CCN-1copy with caching).

To better analyse the protocol behaviour, we also investigate the percentage of successful requested content divided per community. Figure \ref{fig:deliveryRate_LocalForReq_scen3comm} breaks down the overall delivery rate in two components: the delivery rate achieved for the requests of contents within the home community, and the delivery rate achieved for the requests of contents within the foreign communities. In this figure,  the maximum delivery rate (i.e., 1) is achieved when all the 20 requests within the corresponding community are satisfied. As Figure \ref{fig:deliveryRate_LocalForReq_scen3comm:deliveryRate_LocalReq} highlights, MobCCN/noReTrans is able to deliver almost all the requested contents in the home community, and achieves a delivery rate around 0.4 for the foreign communities (see Figure \ref{fig:deliveryRate_LocalForReq_scen3comm:deliveryRate_ForReq}), regardless the use or not of caching mechanism. Conversely, the delivery rate of Epidemic-CCN-1copy/noReTrans is lower than MobCCN/noReTrans. Specifically, in case of requests within the home community, Epidemic-CCN-1copy/noReTrans with no cache achieves a delivery rate around 0.4, while with cache the delivery rate is around 0.55. The performance of Epidemic-CCN-1copy/noReTrans drastically degrades in Figure \ref{fig:deliveryRate_LocalForReq_scen3comm:deliveryRate_ForReq}, where only 1 out of 20 content requests is successfully delivered. This is mainly due to the fact that a unique copy of each request for each content is allowed to circulate in the network, and such request crosses several hops before reaching first the traveller and then the content provider outside the home community, mainly because the next forwarder is searched on a random basis. On the contrary, in the same condition, MobCCN/noReTrans observes the contact patters and forwards only to the next hop with the highest utility, hence achieving higher delivery rates. When more than one copy for each request is allowed to circulate in the network, the performance of Epidemic increases also for the foreign community (see the Epidemic-CCN-1copy bar in Figure \ref{fig:deliveryRate_LocalForReq_scen3comm:deliveryRate_ForReq}). Finally, when the Interests are replicated on all the encountered nodes, the performance of Epidemic reaches the highest value (i.e., 0.5) both for home and foreign communities.

\begin{figure}[] 
\centering
    \subfloat[\label{fig:e2e_LocalForReq_scen3comm:e2e_LocalReq}]{%
      \includegraphics[trim={0cm 0cm 0cm 0cm},clip,angle=-90,width=0.5\textwidth]{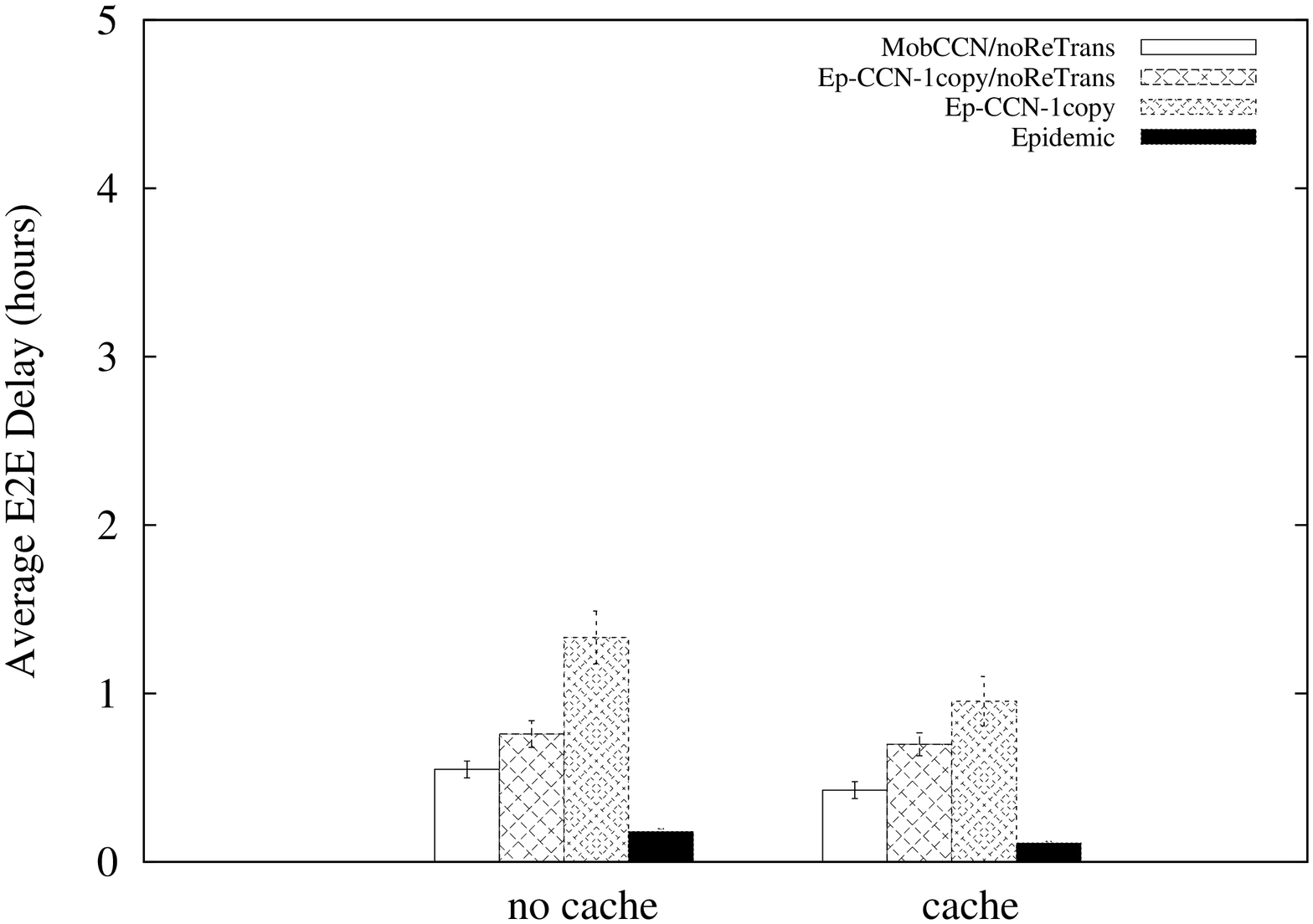}
    }
%
%
    \subfloat[\label{fig:e2e_LocalForReq_scen3comm:e2e_ForReq}]{%
      \includegraphics[trim={0cm 0cm 0cm 0cm},clip,angle=-90,width=0.5\textwidth]{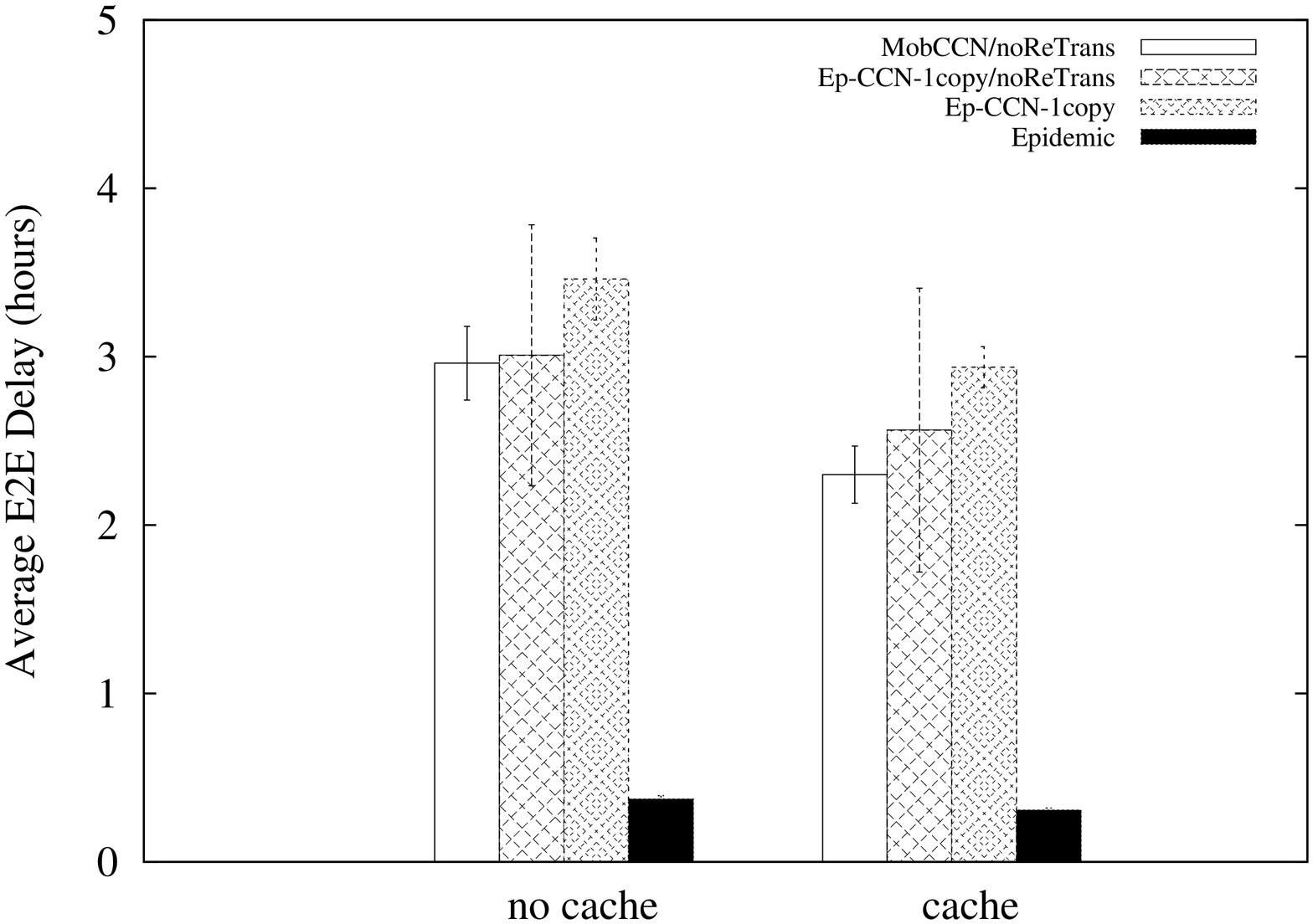}
      }
\caption{Breakdown of the end-to-end delay: home community (\ref{fig:e2e_LocalForReq_scen3comm:e2e_LocalReq}), and foreign communities. (\ref{fig:e2e_LocalForReq_scen3comm:e2e_ForReq}).}
\label{fig:e2e_LocalForReq_scen3comm}
\end{figure}

Figure \ref{fig:e2e_LocalForReq_scen3comm} shows the end-to-end delays experienced by the four protocols divided per community. With respect to this performance index, ideal Epidemic experiences the minimal end-to-end delays as expected. As far as the other protocols, MobCCN/noReTrans outperforms both Epidemic-CCN-1copy/noReTrans and Epidemic-CCN-1copy experiencing lower delays for retrieving contents inside and outside the home community. Obviously, the end-to-end delays of the foreign communities are higher than the values within the home community (about three times), and this is due to the longer paths observed by Data packets in foreign communities (as we will see in the next section). A final consideration concerns the data caching that, when used, has the effect to reduce the experienced delays.

This first part of the analysis confirms that MobCCN/noReTrans has good delivery performance indices, so that to achieve similar performance, at least Epidemic-CCN-1copy with caching needs to be used. In the next section we show how much this should be paid in terms of network resource consumption. 

\begin{figure}[]
	\centering
	\includegraphics[trim={0cm 0cm 0cm 0cm}, clip, angle=0, scale=0.3]{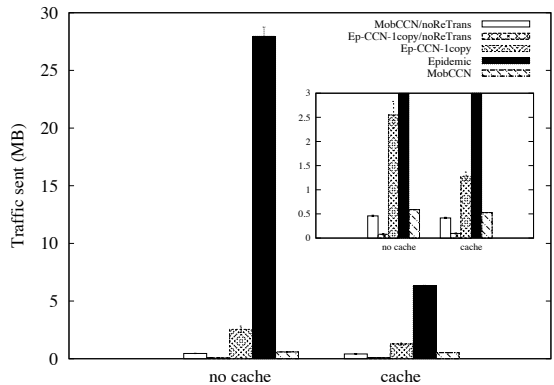}
	\caption{Total traffic sent for forwarding Interest and Data packets}
	\label{fig:trafficSent_scen3comm}
\end{figure}

\subsubsection{Resource consumption}
\label{sub:resource-consumption-scenarioB}

As explained in Section \ref{sub:investigated-scenarios}, in this scenario we analyse separately the two main components of  the total traffic sent, i.e., the forwarding traffic due to the transmission of Interest and Data packets, and the control traffic due to the transmission of routing messages.  
Figure \ref{fig:trafficSent_scen3comm} shows the total traffic due to the forwarding of Interest and Data packets. As apparent from the figure, MobCCN produces a higher network overhead with respect to MobCCN/noReTrans. This result, if added to no substantial increase in delivery performance, is a further confirmation that the retransmission mechanism is not so beneficial for MobCCN but it represents only an addition cost. 
Focusing on the MobCCN/noReTrans performance, the figure highlights that it significantly reduces the traffic transmitted over the network.  Specifically, it cuts the number of Interest and Data packets by over 50 times with respect to the ideal Epidemic in case of "no cache", and by over 10 times when data caching is enabled. MobCCN/noReTrans also generates less traffic than Epidemic-CCN-1copy. On the contrary, when the retransmission mechanism is disabled in Epidemic-CCN-1copy, the latter generates less traffic than MobCCN/noReTrans but this result is paid with an extremely low delivery rate (see Figure \ref{fig:deliveryRate_scen3comm}). MobCCN/noReTrans can therefore considerably limit network resource consumption, and consequently it represents a good compromise because it achieves high delivery rates with moderate forwarding traffic. 

Concerning the second traffic component, i.e., the control traffic, it is worth pointing out that Epidemic-like protocols do not produce any additional routing traffic as nodes relies on random searches for detecting the next forwarder.  Conversely, MobCCN generates additional control traffic as it relies on an exchange of routing packets to detect the most suitable next hop. Specifically, in this scenario we measure about 3.5 byte/sec of routing traffic per node, which is a reasonable value to pay off to allow a more efficient forwarding of Interest packets. It is important to point out that the routing traffic introduced by MobCCN is independent of the number of content requests, but it relies on number of nodes and their mobility patterns. Note also that at the time of writing MobCCN does not implement any optimized mechanisms to reduce such routing overhead, but efficient mechanisms such compression techniques might be designed to further limit the information exchanged by nodes to populate FIBs.

\begin{figure}[h]
	\centering
	\includegraphics[trim={0cm 0cm 0cm 0cm}, clip, angle=-90, scale=0.3]{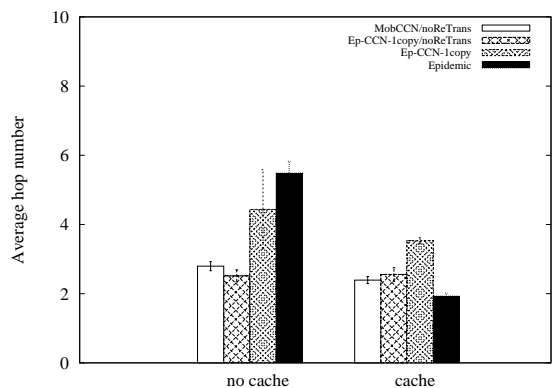}
	\caption{Number of hops taken by Data packets.}
	\label{fig:nHop_scen3comm}
\end{figure}
%

Another interesting performance metric is the number of hops taken by Data packets, which allows to measure how many nodes are involved in the forwarding process. The results are shown in Figure \ref{fig:nHop_scen3comm}. As far as this index, MobCCN/noReTrans performance is equivalent to the corresponding Epidemic version without the retransmission mechanism, with about 2 hops on average. In addition, the number of hops in MobCCN/noReTrans is significantly lower than both Epidemic-CCN-1copy and ideal Epidemic, when no caching mechanism is enabled. This confirms the ability of MobCCN to achieve good delivery rate performance, not only by greatly reducing the traffic sent, but also the number of network resources involved (i.e., nodes). The use of data cache has no effect on MobCCN/noReTrans and Epidemic-CCN-1copy/noReTrans, while it produces a decrease in terms of hop number for Epidemic-CCN-1copy and ideal Epidemic, thus limiting the number of nodes traversed by Data packets.

\begin{figure}[h] 
\centering
    \subfloat[\label{fig:nHop_LocalForReq_scen3comm:nHop_LocalReq}]{%
      \includegraphics[trim={0cm 0cm 0cm 0cm},clip,angle=-90,width=0.5\textwidth]{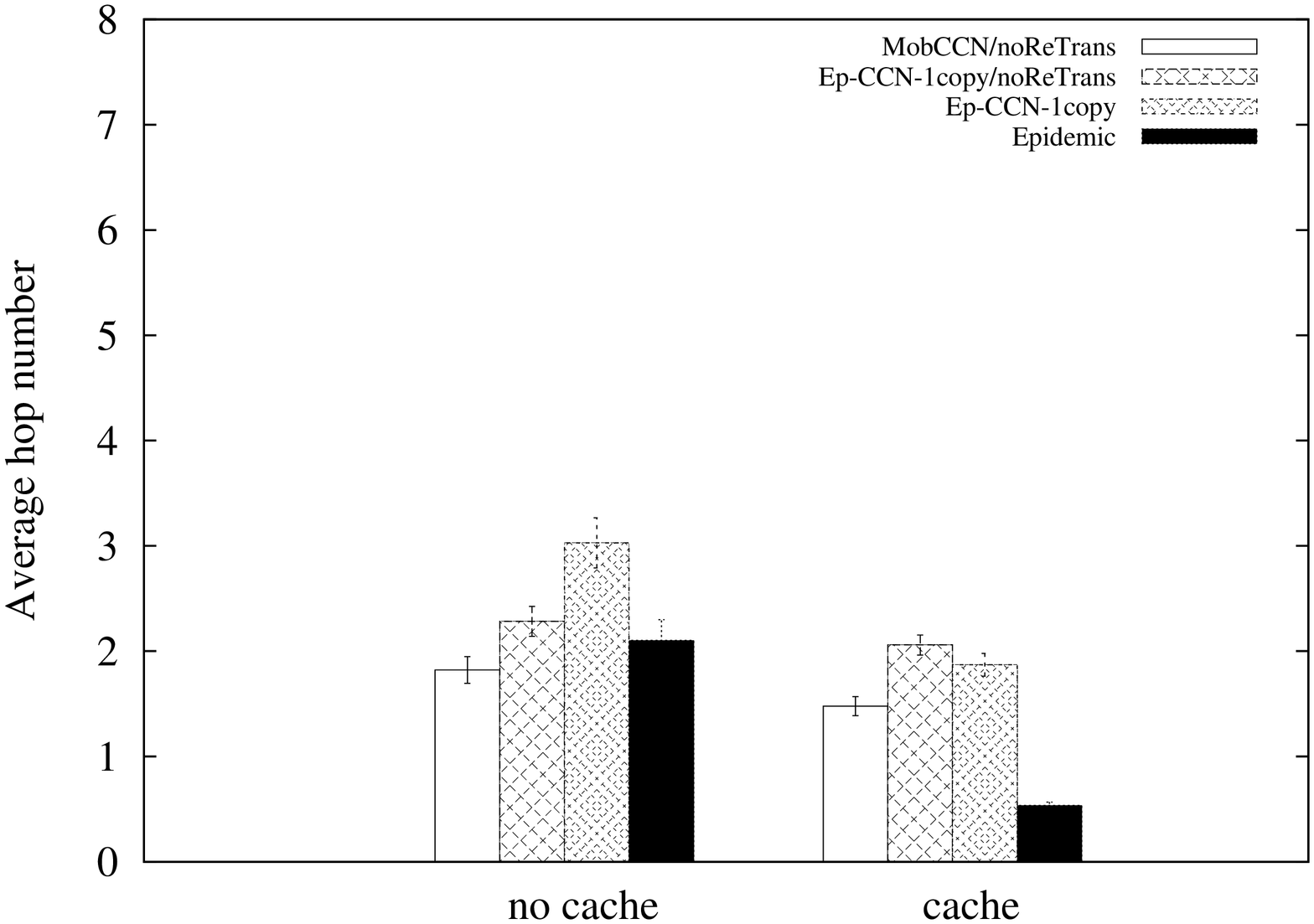}
    }
%
%
    \subfloat[\label{fig:nHop_LocalForReq_scen3comm:nHop_ForReq}]{%
      \includegraphics[trim={0cm 0cm 0cm 0cm},clip,angle=-90,width=0.5\textwidth]{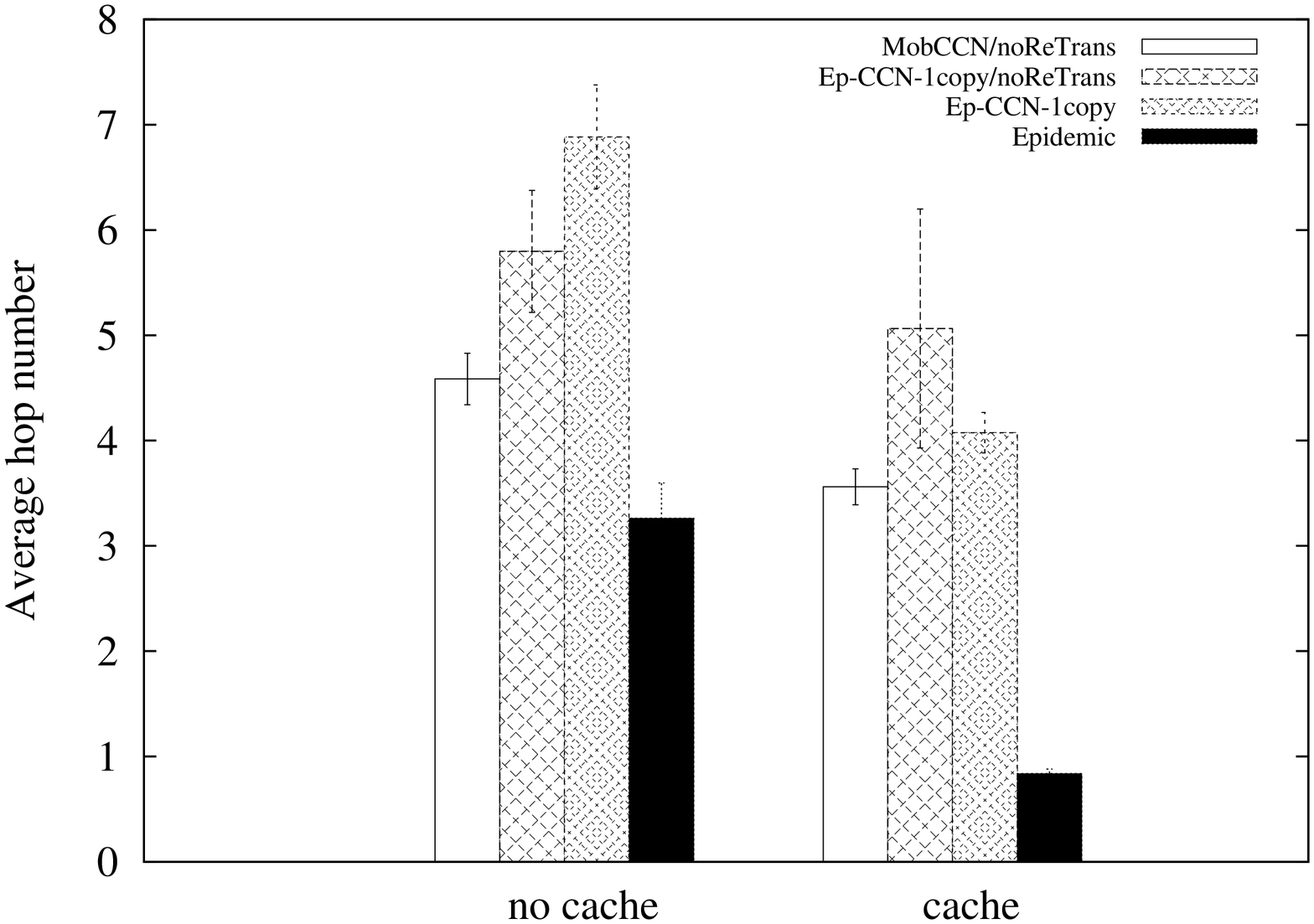}
      }
\caption{Breakdown of number of hop: home community (\ref{fig:nHop_LocalForReq_scen3comm:nHop_LocalReq}), and foreign communities (\ref{fig:nHop_LocalForReq_scen3comm:nHop_ForReq}).}
\label{fig:nHop_LocalForReq_scen3comm}
\end{figure}

As did before for the delivery rate and the end-to-end delay, we break down the number of hops taken by Data packets per community. Figure \ref{fig:nHop_LocalForReq_scen3comm} shows results for the home community and for the foreign communities. Overall, paths followed by Data packets for contents within the home community are shorter than outside it for all the four protocols. As far as the individual protocol performance, in Figure \ref{fig:nHop_LocalForReq_scen3comm:nHop_LocalReq} MobCCN/noReTrans paths are slightly shorter than the other three protocols with no cache (i.e., less than 2 hops), while the presence of local cache favours the classic Epidemic reducing to 0.5 the average number of hops. Having a hop number lower than 1 means that a large part of contents is already cached on the consumers when requested. Concerning the average number of hops for the foreign community, MobCCN/noReTrans paths are lower than the corresponding value of Epidemic value (i.e., Epidemic-CCN-1copy/noReTrans), but obviously they are longer than those in ideal Epidemic, which exploits the huge amount of Data replicas to reduce the path lengths.  

\begin{figure}[]
	\centering
	\includegraphics[trim={0cm 0cm 0cm 0cm},clip,angle=-90, scale=0.3]{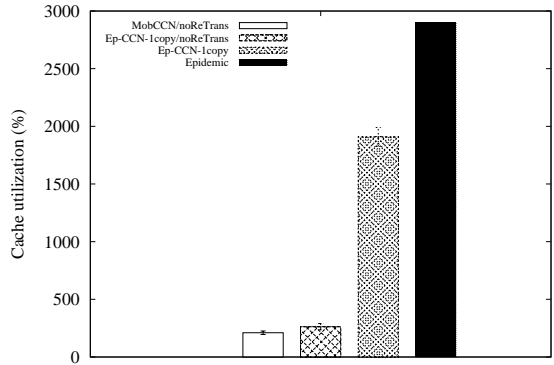}
	\caption{Increase of the cache size}
	\label{fig:cacheSize_scen3comm}
\end{figure}

 Figure \ref{fig:cacheSize_scen3comm} depicts the percentage increase of cache utilization with respect to its initial value according to eq. \ref{eq:cache-utilization}. In this figure, the minimum value 0\% means that at the end of the simulation no nodes - excluding the initial providers - have cached data, while the maximum value for this scenario is 2900\%, i.e., all the 30 nodes have cached all the available contents. We can observe that MobCCN/noReTrans and Epidemic-CCN-1copy/noReTrans make a moderate use of the local cache, with a 250\% increase of the cache utilization with respect to its initial size. This means that about 10-12\%\footnote{This percentage is computed as the total number of contents cached with respect to the maximum value, which is 1800 for this scenario, i.e., all the 60 contents are cached on all the 30 nodes.}of existing contents are cached on nodes at the end of simulation. Conversely, the percentage of cache utilization used by the other two protocols is much higher. Specifically, in Epidemic-CCN-1copy it is around 2000\% meaning that about 67\% of contents are replicated, while in ideal Epidemic it is 2900\%, that is, all the contents are cached on all the nodes consuming considerably device resources. This is a further demonstration of the advantages of MobCCN: its ability to achieve good delivery performance by limiting both the overhead and device resource consumption.

\begin{figure}[tb]
	\centering
	\includegraphics[trim={0cm 0cm 0cm 0cm},clip,angle=-90, scale=0.3]{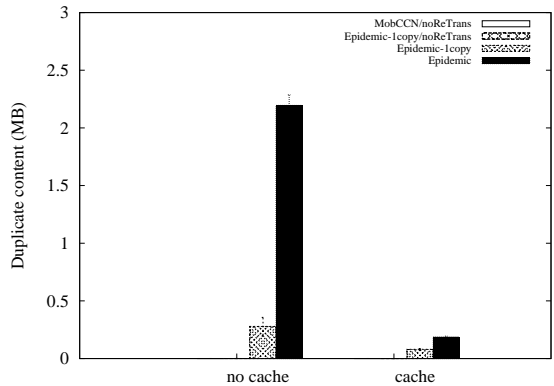}
	\caption{Duplicate content on consumers}
	\label{fig:duplicateContent_scen3comm}
\end{figure}

The final metric we analyse is the number of duplicate contents received by consumers. As apparent from Figure \ref{fig:duplicateContent_scen3comm}, MobCCN/noReTrans and Epidemic-CCN-1copy/noReTrans do not receive any useless copies of requested content. Indeed, when the retransmission mechanism is disabled, each node forwards only one Interest packet for each required content, receiving only one Data packet and no further copies. Conversely, when the retransmission mechanism is active, a number of useless copies are received by consumers because each node treats each Interest packet independently of the others, that is, more than one Interest packet for the same content might circulate over the network, and more than one Data packet might be received by each consumer. This explains the presence of duplicate packets for Epidemic-CCN-1copy. Things worsen in the case of ideal Epidemic where a large number of duplicate content is received by consumers due to an uncontrolled replication of Interest and Data packets transmitted over the network, resulting in a huge waste of network and device resources. The presence of caching mechanism partially limits the percentage of useless copies received as the required content can be retrieved also from the local cache, avoiding forwarding Interests further.
%

\section{Conclusions}
\label{sec:conclusions}

In this paper we have proposed MobCCN, which is a data-centric protocol for IoT environments populated also by mobile nodes forming opportunistic networks. Specifically, the mixing of typical IoT environments (small nodes generating data about the physical environments and receiving commands to act upon it) and mobile opportunistic networking is one of the original traits of MobCCN. This type of heterogeneous mobile networks are expected to be more and more important in the IoT ecosystem, not only due to the exponentially increasing number of mobile devices, but also because they can support agile network solutions, whereby data access does not necessarily depend on the existence of an extremely capillary high-capacity wireless infrastructure, whose capacity needs to constantly match the traffic demand of mobile nodes.

Starting from these assumptions, MobCCN is ICN-compliant, and supports the integration of this type of IoT environments with more conventional fixed ICN Internet infrastructures. Furthermore, it supports CCN-compliant access to data also locally, i.e., among the mobile and IoT devices deployed in a given physical area. This is also a configuration that is gaining more and more attention, as it naturally follows from the typical locality of human behaviour, and the emerging trend of decentralisation of network operations towards the edge.

Contrarily to most of the literature on ICN protocols for mobile environments, MobCCN includes a routing protocol through which nodes populate their FIBs, as a way to guide the propagation of Interest packets. This avoids to use resource inefficient schemes to propagate Interest packets which are often proposed in the literature, such as Epidemic routing. Instead, MobCCN routing exploits the concept of utility of nodes for a given type of content, which is a measure of how frequent a given node ``encounters'' that type of content in other nodes it meets. FIBs are thus set so that Interest packets propagate along increasing gradients of utility, until they reach a node storing the requested data.

Performance results show the efficiency and effectiveness of MobCCN. Specifically, in cases of homogeneous mobilities (i.e., where nodes encounter each other according to identical stochastic processes), we compared it against two opposite cases, both based on variations of Epidemic routing. In the first case, we compare MobCCN against a version of Epidemic that guarantees the minimum end-to-end delay, at the cost of uncontrolled replication of Interest and Data packets. In the second case, we compare MobCCN against a version of Epidemic that uses only one copy of Interest packet, and forwards Data packets along a single path route (using the standard breadcrumbs established by the corresponding Interest packets). With respect to the first benchmark, MobCCN achieves a drastic reduction of the generated traffic, with a reasonable increase of end-to-end delay, without reducing the delivery ratio. With respect to the second benchmark, MobCCN achieves a significant improvement of both end-to-end delay and delivery ratio, with a modest increase of the network traffic. In cases of heterogeneous mobility, the advantage of MobCCN over Epidemic-style routing are even higher. Specifically, in networks composed of different social communities where nodes makes requests also of contents of foreign communities,  MobCCN is able to achieve high delivery rates with lower end-to-end delays and shorter paths even for requests outside their "home" community. At the same time, it is also able to keep the consumption of network and device resources moderate. Conversely, to achieve the same MobCCN performance, Epidemic-style protocols have to generate huge network traffic and/or excessively waste the local resources of the devices. Finally, note that the retransmission mechanism implemented by MobCCN contributes to an overall increase of costs only, but it is not so beneficial in terms of delivery performance. 

In conclusion, we can claim that the fundamental idea underpinning the definition of MobCCN proves to be efficient and effective. Specifically, ``investing'' some traffic to build FIBs makes sense, if this is done in a way that does not generate excessive overhead, but still provides correct information to discriminate among different nodes as potential forwarders. Finally, to the best of our knowledge, MobCCN is one of the few fully ICN-compliant protocols for IoT environments in presence of mobile nodes. This makes it particularly attractive, due to the increasing diffusion of ICN as one of the reference networking paradigms for the Internet.

\section*{Acknowledgments}
This work is partly funded by the EC under the H2020
REPLICATE (691735), SoBigData (654024) and AUTOWARE
(723909) projects.

\section*{References}
\bibliography{comcom_5663}

\end{document}